\documentclass{emulateapj}

\pdfoutput=1

\usepackage{color}
\usepackage{graphicx}

\usepackage{natbib}
\slugcomment{Accepted for Publication in the Astrophysical Journal October 8, 2015}

\shorttitle{LEECH: NO UMa}
\shortauthors{J. E. Schlieder et al.}

\begin{document}


\title{The LEECH Exoplanet Imaging Survey: Orbit and Component Masses of the Intermediate Age, Late-Type Binary NO UMa\footnotemark[*,]\footnotemark[$\dagger$]}


\author{Joshua E. Schlieder\altaffilmark{1, 2}, Andrew J. Skemer\altaffilmark{3}, Anne-Lise Maire\altaffilmark{4}, Silvano Desidera\altaffilmark{4}, Philip Hinz\altaffilmark{3}, Michael F. Skrutskie\altaffilmark{5}, Jarron Leisenring\altaffilmark{3}, Vanessa Bailey\altaffilmark{6}, Denis Defr\`ere\altaffilmark{3}, Simone Esposito\altaffilmark{7}, Klaus G. Strassmeier\altaffilmark{8},  Michael Weber\altaffilmark{8},  Beth A. Biller\altaffilmark{9, 2}, Micka\"el Bonnefoy\altaffilmark{10, 2}, Esther Buenzli\altaffilmark{2}, Laird M. Close\altaffilmark{3}, Justin R. Crepp\altaffilmark{11},  Josh A. Eisner\altaffilmark{3}, Karl-Heinz Hofmann\altaffilmark{12}, Thomas Henning\altaffilmark{2}, Katie M. Morzinski\altaffilmark{3}, Dieter Schertl\altaffilmark{12}, Gerd Weigelt\altaffilmark{12}, Charles E. Woodward\altaffilmark{13}}

\footnotetext[*]{The LBT is an international collaboration among institutions in the United States, Italy and Germany. LBT Corporation partners are: The University of Arizona on behalf of the Arizona university system; Istituto Nazionale di Astrofisica, Italy; LBT Beteiligungsgesellschaft, Germany, representing the Max-Planck Society, the Astrophysical Institute Potsdam, and Heidelberg University; The Ohio State University, and The Research Corporation, on behalf of The University of Notre Dame, University of Minnesota and University of Virginia.}
\footnotetext[$\dagger$]{Based on data obtained with the STELLA robotic telescope in Tenerife, an AIP facility jointly operated by AIP and IAC.} 
\altaffiltext{1}{NASA Postdoctoral Program Fellow, NASA Ames Research Center, Space Science and Astrobiology Division, MS 245-6, Moffett Field, CA 94035, USA}
\altaffiltext{2}{Max-Planck-Institut f\"ur Astronomie, K\"onigstuhl 17, 69117, Heidelberg, Germany}
\altaffiltext{3}{Steward Observatory, Department of Astronomy, University of Arizona, 933 N. Cherry Ave, Tucson, AZ 85721, USA}
\altaffiltext{4}{INAF - Osservatorio Astronomico di Padova, Vicolo dell'Osservatorio 5, 35122, Padova, Italy}
\altaffiltext{5}{Department of Astronomy, University of Virginia, Charlottesville, VA, 22904, USA}
\altaffiltext{6}{Kavli Institute for Particle Astrophysics and Cosmology, Stanford University, Stanford, CA 94305, USA}
\altaffiltext{7}{INAF - Osservatorio Astrofisico di Arcetri, Largo E. Fermi 5, 50125, Firenze, Italy}
\altaffiltext{8}{Leibniz-Institut f\"ur Astrophysik Potsdam (AIP), An der Sternwarte 16, 14482, Potsdam, Germany}
\altaffiltext{9}{Institute for Astronomy, University of Edinburgh, Blackford Hill, Edinburgh EH9 3HJ, UK}
\altaffiltext{10}{Universit\'e Grenoble Alpes, IPAG, 38000, Grenoble, 38000, Grenoble; CNRS, IPAG, 38000 Grenoble, France}
\altaffiltext{11}{Department of Physics, University of Notre Dame, 225 Nieuwland Science Hall, Notre Dame, IN, 46556, USA}
\altaffiltext{12}{Max-Planck-Institut f\"ur Radioastronomie, Auf dem H\"ugel 69, 53121, Bonn, Germany}
\altaffiltext{13}{Minnesota Institute for Astrophysics, University of Minnesota, 116 Church Street, SE, Minneapolis, MN, 55455, USA}
\email{joshua.e.schlieder@nasa.gov}






\begin{abstract}

{We present high-resolution Large Binocular Telescope LBTI/LMIRcam images of the spectroscopic and astrometric binary NO UMa obtained as part of the LBTI Exozodi Exoplanet Common Hunt (LEECH) exoplanet imaging survey. Our $H, K_s$, and $L^{\prime}$ band observations resolve the system at angular separations $<0\farcs09$. The components exhibit significant orbital motion over a span of $\sim$7 months. We combine our imaging data with archival images, published speckle interferometry measurements, and existing spectroscopic velocity data to solve the full orbital solution and estimate component masses. The masses of the K2.0$\pm$0.5 primary and K6.5$\pm$0.5 secondary are $0.83\pm0.02$ M$_{\odot}$ and $0.64\pm0.02$ M$_{\odot}$, respectively. We also derive a system distance of $d = 25.87 \pm 0.02$ pc and revise the Galactic kinematics of NO UMa. Our revised Galactic kinematics confirm NO UMa as a nuclear member of the $\sim$500 Myr old Ursa Major moving group and it is thus a mass and age benchmark. We compare the masses of the NO UMa binary components to those predicted by five sets of stellar evolution models at the age of the Ursa Major group. We find excellent agreement 
between our measured masses and model predictions with little systematic scatter between the models. NO UMa joins the short list of nearby, bright, late-type binaries having
known ages and fully characterized orbits.}
\end{abstract}






\section{Introduction}
\setcounter{footnote}{0}

Multiple star systems are a natural outcome of the star-formation process. Thus, stars in binaries
and higher order systems are prevalent in the solar neighborhood. \cite{RAGHAVAN10} surveyed
more than 450 solar-type stars ($M_*\approx0.7 - 1.1~M_{\odot}$, mid-K to late-F spectral types) 
in a 25 pc volume around the Sun and found a multiplicity fraction of $41\pm3$\%. The majority 
of these multiples are binaries \citep{DUCHENE13}. Much less frequent however, are multiple systems
amenable to detailed orbit characterization. Distance, separation, and 
mass must all be favorable to make the measurement of orbital parameters feasible on reasonable
 timescales using both spectroscopic and astrometric monitoring.  Analyses of such systems 
 provide precise estimates of parameters such as period, eccentricity, inclination, and most 
 critically, component masses. If these systems have accurately determined ages, they 
 act as benchmarks for understanding the evolution of fundamental stellar parameters and allow 
 the calibration of widely used stellar evolution models.     
 
One such system in the solar neighborhood is NO UMa (HIP 61100, HD 109011, GJ 1160). 
NO UMa is a pair of K-type dwarfs at a distance of $d = 25.87 \pm 0.02$ pc (see \S~\ref{orbit_fit}). 
The system was observed during the CORAVEL radial velocity (RV)
survey \citep{BARANNE79, DUQUENNOY91} where its spectroscopic binarity was discovered. 
At the time, no CORAVEL RV curve or spectroscopic orbit parameters for NO UMa were presented in the 
literature, although the statistical studies of K-type binaries in \citet{MAYOR1992} and 
\citet{HALBWACHS00} from the CORAVEL survey presumably included the system.
 \citet{ARENOU00} presented the first orbit solution, including preliminary component masses, 
 using the CORAVEL RV data and Intermediate Astrometric Data from $Hipparcos$ catalog \citep{PERRYMAN97}.

Further follow-up by \cite{STRASSMEIER00} revealed chromospheric activity and Li absorption, 
indications of a relatively young age. The first 
spectroscopic orbit parameters were provided by \citet{HALBWACHS03} and an independent 
astrometric orbit from the $Hipparcos$ data was presented in \cite{GOLDIN07}. 
In a subsequent paper, \cite{STRASSMEIER12} presented updated RV curves for 
both components of the binary, provided a spectroscopic orbit solution, and updated
fundamental and spectroscopic parameters of each component. The orbit parameters estimated 
in these studies and the $Hipparcos$ distance indicated that the components of the NO UMa 
system may be resolvable at angular separations $\lesssim0\farcs1$; feasible 
with modern adaptive optics (AO) systems on large aperture telescopes. Additionally, \cite{KING03}
proposed the star as a nuclear member of the Ursa Major moving group (or cluster), a $\sim$500 Myr old
group of coeval stars with common Galactic kinematics. 
 
\begin{table*}[!t]
\begin{center}
\caption{Summary of NO UMa physical properties \label{tab1}}
\begin{tabular}{lcccc}
\tableline\tableline
 & NO UMa & NO UMa A & NO UMa B & Reference \\
\tableline
$\alpha_{(J2000)}$ ($^{\circ}$) & 187.828876   & \dots  & \dots  & 1 \\ 
$\delta_{(J2000)}$ ($^{\circ}$) & +55.118858	   &  \dots  & \dots  & 1 \\
$\mu_{\alpha}$ (mas yr$^{-1}$) & $107.08\pm1.20$  & \dots  & \dots & 2 \\
$\mu_{\delta}$ (mas yr$^{-1}$) & $0.38\pm1.22$  & \dots  & \dots  & 2 \\
d (pc) &  $25.87\pm0.02$ & \dots & \dots  &  8 \\
systemic RV (km s$^{-1}$) & -$9.873\pm0.007$   & \dots & \dots   & 8 \\
$v$sin$i$ (km s$^{-1}$) & \dots & $5\pm1$ & $6\pm1$ & 3 \\
$U$(mag) &  $9.70\pm0.03$  &  \dots  &  \dots   &  4  \\
$B$(mag) &  $9.05\pm0.03$  &  \dots  &  \dots   &  4  \\
$V$(mag) &  $8.13\pm0.03$  &  \dots  &  \dots   &  4  \\
$B_T$(mag) &  $9.29\pm0.02$  &  \dots  &  \dots   &  5  \\
$V_T$(mag) &  $8.21\pm0.01$  &  \dots  &  \dots   &  5  \\
$J$ (mag) &  $6.32\pm0.03$  &  \dots  &  \dots   &  1  \\
$H$ (mag) & $5.81\pm0.03$  & $6.27\pm0.10$ & $6.96\pm0.12$  & 1,8 \\
$K_{s}$ (mag) &  $5.66\pm0.02$  &  $6.12\pm0.03$   & $6.83\pm0.04$   &  1,8 \\
$L'$ (mag) & $\sim$$5.64\pm0.15$  & $6.09\pm0.15$  & $6.82\pm0.15$ & 6,8 \\
$\mathrm{T_{eff}}$ (K)   &   \dots    &   $5010\pm50$  & $4140\pm30$  & 8 \\
Spectral Type   &  K2Ve $\pm$ 1   &   K2.0V$\pm$0.5  & K6.5V$\pm$0.5  & 3,8 \\
$\mathrm{log(L/L_{\odot})}$ (dex) &  \dots & -0.49$\pm$0.03  &  -0.97$\pm$0.02    &  8 \\
Mass ($\mathrm{M_{\odot}}$)  &  $\mathrm{1.47\pm0.03}$  & $\mathrm{0.83\pm0.02}$  &  $\mathrm{0.64\pm0.02}$   &    8  \\
Age (Myr) & 500$\pm$100 & \dots  & \dots   & 7 \\ 
\tableline
\label{properties_table}
\end{tabular}
\tablecomments{1 - \citet{CUTRI03}; 2 - \citet{ARENOU00}; 3 - \citet{STRASSMEIER12} 4 - \citet{MERMILLIOD94}; 5 - \citet{HOG00}; 6 - \citet{CUTRI13, WRIGHT10}; 7 - \citet{KING03, BRANDT15}; 8 - This work}
\end{center}
\end{table*}

Thus, NO UMa is an attractive target for high angular resolution, high-contrast AO imaging, not only to 
resolve the binary components, but also to search for low-mass companions. Stellar binaries are typically excluded 
from exoplanet imaging surveys, however, NO UMa's component separation is small enough ($\sim$2-3 AU) 
that circumbinary companions on wide orbits are not dynamically unstable \citep[][and references therein]{THALMANN14}. Although 
challenging for planet formation theory \citep{KLEY14}, recent work has revealed circumbinary disks with the potential 
to form planets \citep[e.g.][]{RAPSON15, TANG14, DUTREY14}, a few binaries with directly imaged, circumbinary, planet/brown dwarf companions 
\citep{KRAUS14, DELORME13}, and  numerous circumbinary planets in transit \citep[][and references therein]{WELSH15}. For these reasons, 
we included NO UMa as a target in the LEECH exoplanet imaging survey \citep{SKEMER14SPIE} and succeeded in resolving the individual
components.

In this work, we describe the derivation of a full set of orbital parameters for NO UMa to provide
component masses, estimate the fundamental parameters of each component using our 
resolved photometry, revise the system's Galactic kinematics using new measurements from our orbit fit,
and compare our measured masses to model predictions.
In \S 2 we summarize the LEECH program and NO UMa's inclusion as a 
target. In \S 3 we provide details on the available fundamental
properties and orbital parameters of NO UMa. \S 4 
describes our AO imaging of the target, an archival imaging data set, and 
the data reduction.  In \S 5 we describe the imaging analyses and results. 
We combine our astrometric measurements with existing data to solve the complete orbit of the binary 
and improve component mass constraints by a factor $\gtrsim$6 in \S 6. We present in \S 7 fundamental parameters 
and revised kinematics of the binary components and compare the estimated 
component masses to those predicted by theoretical evolution models. \S 8 provides a summary.

\section{The LEECH Exoplanet Imaging Survey}

The LBTI Exozodi Exoplanet Common Hunt (LEECH) is a multi-national collaboration using the Large
Binocular Telescope (LBT) coupled with the dual deformable secondary LBT AO system 
\citep[FLAO, LBTIAO,][]{ESPOSITO10, ESPOSITO11, RICCARDI10, BAILEY14}. LEECH uses the $L/M$-band Infrared Camera 
\citep[LMIRcam,][]{SKRUTSKIE10, LEISENRING12} of the LBT Interferometer \citep[LBTI,][]{HINZ08} 
to conduct the first large scale, exoplanet imaging survey at thermal infra-red (IR) wavelengths ($L^{\prime}$-band, 
$\lambda_{c}$~$\approx$~3.8 $\mu$m) over $\sim$100 nights \citep{SKEMER14SPIE}. LEECH takes
 advantage of two key features of searching for planets in the thermal IR. First, because of strong molecular
 absorption at shorter wavelengths, giant exoplanet fluxes peak between $\sim$4-5 $\mu$m \citep{BURROWS97}. Second, AO 
 systems perform better at longer wavelengths and provide optimal correction \citep{BECKERS93}. Therefore, the LEECH survey 
 is sensitive to older, cooler planets ($\lesssim$1 Gyr, $\lesssim$1000 K) and complements other next generation 
exoplanet surveys searching for younger ($\lesssim$200 Myr), hotter planets in the 
near-IR \cite[GPI, SPHERE, Project 1640,][]{MACINTOSH08, BEUZIT08, HINKLEY11}.

Targets in the LEECH survey span the 
relatively unexplored age range of $\sim$0.1-1 Gyr, a range where LBTI/LMIRcam remains sensitive to both ``hot-start" 
and ``cold-start" planets \citep{SPIEGEL12, MARLEAU14}. The targets are drawn from several samples that include nearby A- and 
B-type stars, very nearby, $\le$1 Gyr old FGK stars, and more than 50 stars in the intermediate 
age Ursa Major moving group. NO UMa is included as a LEECH target in this subsample. The sensitivity and 
utility of the LBTAO coupled with LBTI/LMIRcam has been demonstrated in several studies of 
known, substellar companions and a very low-mass binary  
\citep[Skemer et al.~2015, submitted;][]{SKEMER12, SKEMER14, BONNEFOY14, SCHLIEDER14}. These capabilities also 
led to strong constraints on the possibility of a fifth planet in the HR 8799 planetary system during the 
LEECH survey \citep{MAIRE2015}. Further technical details of the LEECH survey, including 
$H$ and $L^{\prime}$ contrast curves, are provided in \cite{SKEMER14SPIE}.

\section{Known Fundamental and Orbital Properties of NO UMa}\label{sec_props}

NO UMa was identified decades ago as a K2V standard in the Morgan-Keenan (MK) system via 
visual inspection of photographic spectrograms \citep{JOHNSON53}.  This spectral type (SpTy) 
is the integrated type for both components and has changed very little since first proposed. 
Independent determinations in the literature range from $\sim$K1V - K3V \citep[e.g.,][]{YOSS61, HEINZE05}. 
Using available optical photometry for the NO UMa system (Table~\ref{properties_table}) and the 
main-sequence color-temperature conversions of \cite{PECAUT13}\footnote{Throughout this work, we use the expanded table available on 
Eric Mamajek's webpage: http://www.pas.rochester.edu/$\sim$emamajek/
EEM\_dwarf\_UBVIJHK\_colors\_Teff.txt}, we interpolate a median 
SpTy of K2.5$\pm$0.5 using Monte Carlo (MC) methods. We therefore conservatively adopt an integrated system SpTy of 
K2Ve $\pm$ 1 (``e" for emission, see below). 

\cite{ARENOU00} used RV data from a CORAVEL survey of late-type main sequence stars and
Intermediate Astrometric Data from the \emph{Hipparcos} mission to estimate orbital parameters 
for the NO UMa system. The semi-major axis of the \emph{Hipparcos} photocenter was combined 
with the period, eccentricity, mass ratio, and other parameters from the CORAVEL RV curve to place 
constraints on individual component masses. They estimate the primary 
and secondary masses with relative errors of $\sim$25\% and $\sim$17\%, respectively. Their analysis also provided a revised parallax 
and proper motions that were corrected for the motion of the \emph{Hipparcos} photocenter.  

\cite{STRASSMEIER00} obtained high-resolution optical spectroscopy and photometric monitoring 
of NO UMa in their search fzor late-type Doppler-imaging targets. Their Kitt Peak National Observatory 
0.9-m coud\'e feed spectra revealed Ca II H \& K chromospheric emission and weak Li absorption in 
the system with a $34\pm7$ m\AA~equivalent width. Their 
Str\"omgren \emph{y} photometry from the 0.75-m Vienna Observatory automatic photometric 
telescope (APT) ``Wolfgang" provided an estimated period of $\sim$8.3 days. 
The observed activity, Li, and rotation are indicative of an age $\lesssim$625 Myr but $>$125 Myr in 
a $\sim$K2 type star \citep{MAMAJEK08, KING05}.

Following their initial study, \cite{STRASSMEIER12} present dedicated spectroscopic monitoring of 
NO UMa using the 1.2-m STELLA-I telescope and the STELLA Echelle Spectrograph (SES) on Tenerife. 
Their SES data consisted of 129 spectra obtained over 1629 days. They measured individual component 
velocities in each spectrum to generate RV curves and solve the spectroscopic orbit. 
Their high-quality spectra covering the entire orbit allow them to derive a period, time of periastron, eccentricity, 
and systemic RV to $\lesssim$1\% precision. They also reanalyze their APT 
photometric data that consists of 60 observations over 135 days to obtain a new photometric period of 
$8.4\pm0.2$ days. They attribute this period to rotational modulation of the primary. Considering the 
primary's approximately early-K SpTy, this is broadly consistent with its $5\pm1$ km s$^{-1}$ $v$sin$i$.  
\cite{STRASSMEIER12} also derive fundamental and spectroscopic parameters for each component 
using the synthetic spectrum fitting package PARSES \citep{ALLENDE04,JOVANOVIC13}. These include 
effective temperatures of $5030\pm75$ K and $4900\pm150$ K for the primary and secondary, respectively.   
We also note that orbital parameters from the CORAVEL RV data are briefly discussed in \cite{HALBWACHS03} 
and an independent estimate of the orbit from the \emph{Hipparcos} intermediate data is 
presented in \cite{GOLDIN07}.

\cite{BALEGA13} also present speckle interferometry measurements of NO UMa 
from the 6-m BTA telescope at the Special Astrophysical Observatory of the Russian 
Academy of Sciences (SAO-RAS). Their observations span $\sim$4 years from 2002 to 2006 and 
were obtained using filters with $\lambda_{c}$ = 545, 750, or 800 nm.  We describe 
their data in more detail in \S 6.

\begin{figure}[!htb]
\epsscale{1.0}
\plotone{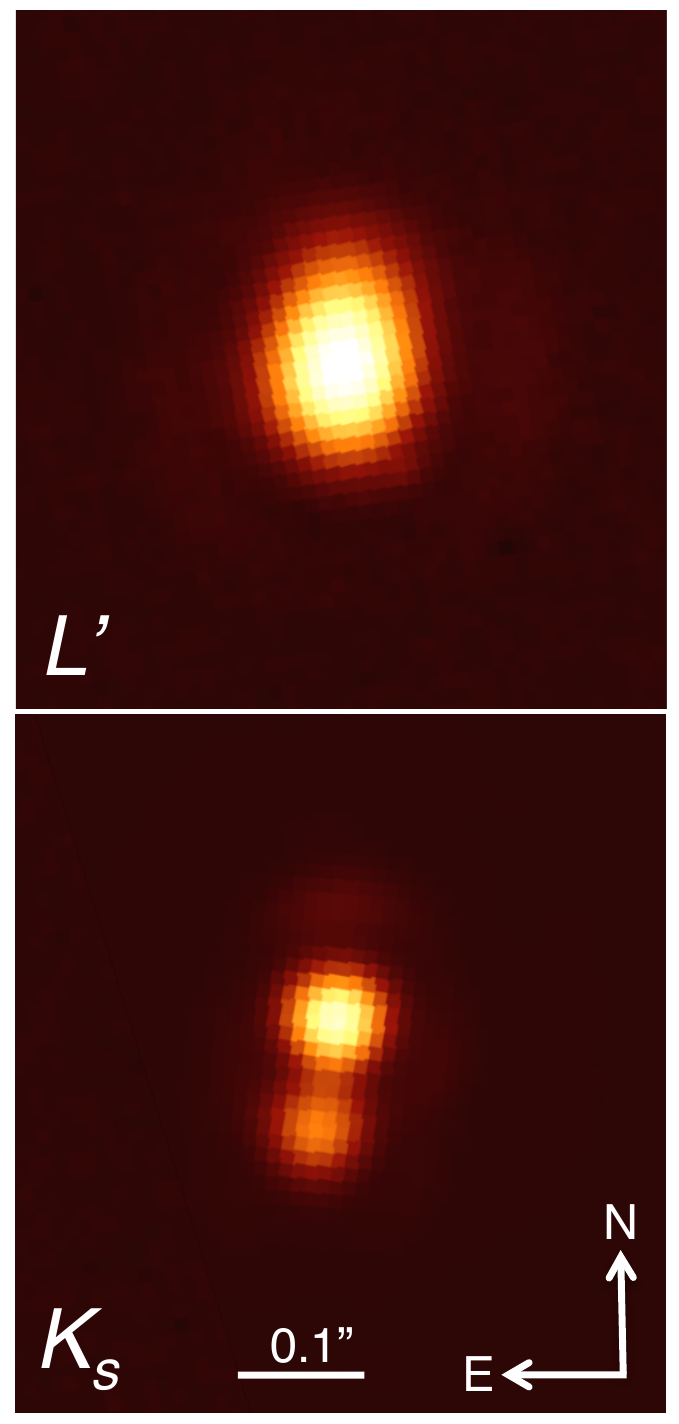}
\caption{LBTI/LMIRcam $L^{\prime}$-band (top) and $K_{s}$-band (bottom) 
images of NO UMa  A and B. The $L^{\prime}$ image was obtained on April 22 2013 UT and the $K_s$ image was 
obtained on December 26 2013 UT.  The binary components are blended at $L^{\prime}$ but well resolved at $K_{s}$.
Our analyses estimate an $L^{\prime}$ separation of $64.7 \pm 0.2$ mas and a $K_{s}$ separation of $86.2 \pm 0.4$ mas.  
The system exhibits $\sim$180$^{\circ}$ of position angle change and $\sim$20 
mas of separation change in only 7 months. \label{figimage}}
\end{figure}

\section{Observations and Data Reduction}

\subsection{LEECH LBTI/LMIRcam Imaging}

NO UMa was observed using LBTI/LMIRcam during two LEECH observing runs in 2013. The LBTI
is located at the bent Gregorian focus of the LBT and does not have a derotator. 
Only the right side of the LBT was used during the observations (the ``DX" side). The LBT AO system 
was driven using NO UMa as a natural guide star.

On April 22 2013 UT we obtained 200 $\times$ 0.495 s exposures of NO UMa with the $L^{\prime}$-band
filter ($\lambda_c = 3.70 \mu$m, $\Delta\lambda = 0.58 \mu$m). The binary was dithered to two positions
in the field of view separated by 4\farcs5. We also observed a star with similar spectral type, HIP 46580 (K3V),
immediately after NO UMa to calibrate the telescope+detector point spread function
(PSF). Our reduction includes corrections for distortion effects, detector bias, sky background, 
and bad pixels followed by frame re-centering via cross-correlation and averaging. The blended components exhibit 
an elongated intensity distribution in the $L^{\prime}$ images (Figure~\ref{figimage}).  We obtained second epoch
 LBTI/LMIRcam images of NO UMa on December 26 UT 2013 in the $H$ 
($\lambda_c = 1.65 \mu$m, $\Delta\lambda = 0.31 \mu$m) and $K_s$-band filters 
($\lambda_c = 2.16 \mu$m, $\Delta\lambda = 0.32 \mu$m). The $H$- and $K_s$-band observations 
each consisted of 100 $\times$ 0.058 s exposures dithered to two positions separated by 
4\farcs5. We followed the same reduction steps for the $H$ and $K_s$ 
frames as for the $L^{\prime}$ frames. The components are
well resolved in both the $H$ and $K_s$-bands. The $K_s$ image is shown in  Figure~\ref{figimage}. 
We observed a photometric calibrator immediately after the observations in both near-IR bands, but the 
PSF of the calibrator was not useful for subsequent analyses due to an issue with the AO
that affected only the calibrator observations (see \S\ref{im_analysis}).

\subsection{Keck-II/NIRC2 Archival Imaging}

\begin{figure}[!htb]
\epsscale{1.0}
\plotone{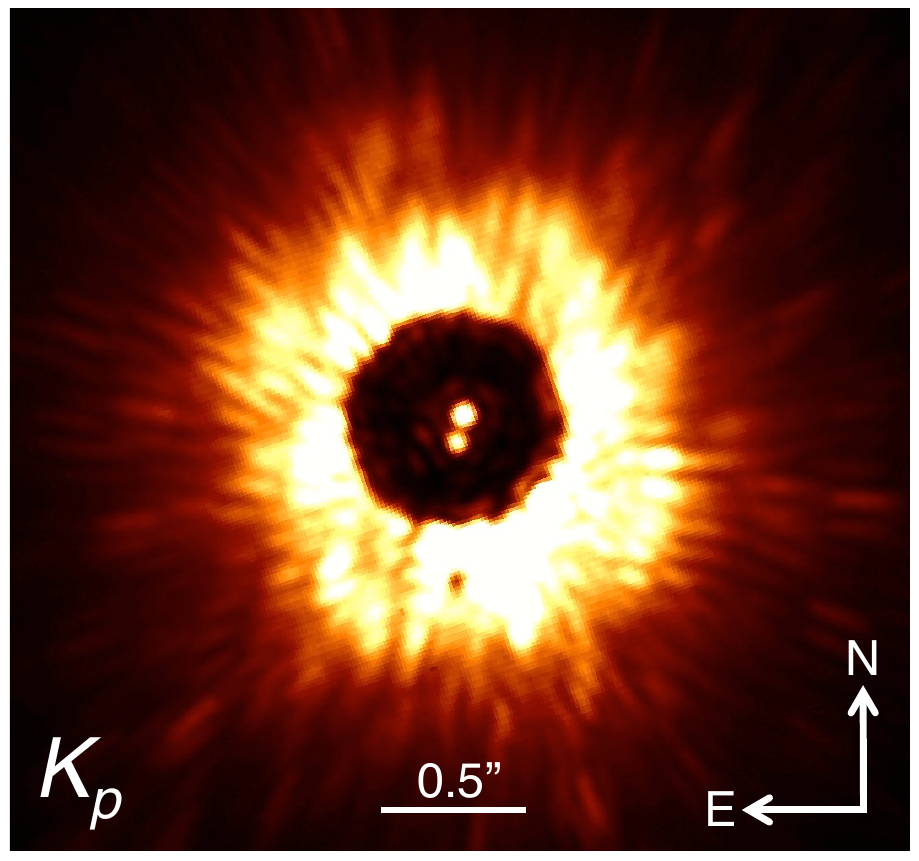}
\caption{Keck/NIRC2 $K_p$-band image of NO UMa A and B. The image was obtained on 
May 27 2010 UT. The binary is clearly resolved behind the 0\farcs6 translucent mask. The components are
separated by $77.7 \pm 0.7$ mas. The position 
angle of the secondary in this earlier epoch is very close to our December 2013 UT 
LBTI/LMIRcam images. The bright halo surrounding the mask is scattered light. \label{keck_image}}
\end{figure}

NO UMa was observed on May 27 2010 UT using Keck-II/Near Infrared Camera 2 (NIRC2) coupled 
with NGS AO\footnote{program ID K319N2, PI Armandoff} \citep{WIZINOWICH00}. The data were 
obtained in the $K_p$-filter ($\lambda_c = 2.124 \mu$m, $\Delta\lambda = 0.351 \mu$m) with the 
narrow camera setting yielding a field-of-view of 10$\farcs$2 $\times$ 10$\farcs$2. Eight frames 
were obtained with NO UMa placed behind the 0$\farcs$6 diameter translucent focal plane mask. 
Three of these images had exposure times of 5.0 s and the remaining had exposure times of 60.0 s. 
The binary is clearly resolved behind the mask in all eight frames (Figure~\ref{keck_image}). Saturated 
images of NO UMa were also obtained at four different dither locations to estimate the sky background. 
The NIRC2 data reduction included cosmic ray and bad pixel removal, dark subtraction, flat fielding, 
and sky subtraction. Optical distortions were corrected using the  NIRC2 distortion solution provided 
by the Keck observatory. 

\section{Image Analyses}\label{im_analysis}

Since the binary is blended in our $L^{\prime}$ images, the component separation 
and flux ratio were calculated by fitting the data with a 
two-star model constructed from the PSF of the nearby standard using Levenberg-Marquardt minimization. 
We find the components have $\Delta L^{\prime} =  0.73 \pm 0.01$ and are separated 
by $6.046\pm0.014$ pixels.   
We converted the measured separation in pixels to mas using the LMIRcam plate 
scale of $10.707 \pm 0.012$ mas pix$^{-1}$ from \citet{MAIRE2015} to 
arrive at an angular separation of $64.7 \pm 0.2$ mas. We calculate the magnitude of the primary using 
$m_1 = m_{12} + 2.5\cdot log_{10}(1+10^{-\Delta m/2.5})$, where $m_1$ is the primary magnitude, $m_{12}$ is the unresolved, integrated
magnitude of both components, and $\Delta m$ is the measured magnitude difference. Since no 
calibrated $L^{\prime}$ photometry is available for NO UMa, we estimate the component photometry using 
the WISE $W1$ magnitude of the system as an approximation for $m_{12}$ in the previous equation. The secondary 
magnitude, $m_2$, is then calculated from $\Delta L^{\prime}$ and $m_1$. The approximate $L^{\prime}$-band system
photometry is provided in Table~\ref{properties_table}. 

We detected no additional companions in the LMIRcam field of view. Since the binary was partially resolved  
in our $L^{\prime}$ images, we did not perform a full deep imaging sequence to search for planetary mass
companions. However, our short integrations were sensitive to circumbinary tertiary companions with $\Delta L^{\prime} =  5$~mag 
at separations $>$0$\farcs5$. At the $\sim$500 Myr age of the system (see \S~\ref{group_membership}), this magnitude ratio and angular separation correspond
to companion masses M~$\gtrsim$~0.1 M$_{\odot}$ \citep{BARAFFE98} at projected separations $\gtrsim$13 AU. 

Issues with the PSF calibrator in the $H$ and $K_s$-bands did not permit analysis of those images 
using the same fitting routine employed for the $L^{\prime}$ data. We modified the procedure to 
allow the PSF to vary as an additional free parameter in the fit minimization. The modified procedure resulted in
a best fit magnitude difference and pixel separation in the $K_s$-band of $\Delta K_s =0.71 \pm 0.02$ mag and 
$8.051 \pm 0.019$ pixels, respectively. In the $H$-band, the new method provides 
$\Delta H = 0.69 \pm 0.06$ and a
separation of $8.02 \pm 0.16$ pixels. The larger uncertainties in the $H$-band are a result of a poorer fit and
larger residuals in the minimization. We conservatively adopt flux ratio and separation errors of 5$\%$ and 2$\%$ in this band. 
We used the LMIRcam plate scale to calculate an $H$-band angular separation of $85.9 \pm 1.7$ mas. When calculating the $K_s$-band separation
in the same way, we find that it does not overlap with the $H$-band separation within 1$\sigma$ uncertainties. To compensate, we add the 
difference between the nominal $H$ and $K_s$ separations in quadrature to the measured $K_s$ 
uncertainty as an extra systematic error to arrive at a final $K_s$ angular separation of $86.2 \pm 0.4$ mas.
Following the same procedure described for the $L^{\prime}$ data, we used the unresolved 
Two Micron All-Sky Survey \citep[$2MASS,$][]{CUTRI03} $H$ and $K_s$-band photometry 
and the measured magnitude differences to calculate the $H$ and $K_s$ component photometry. 
These near-IR magnitudes are listed in Table~\ref{properties_table}.   

Since LMIRcam has no derotator, the $L^{\prime}$, $H$, and $K_s$ images must be re-oriented with true North. 
We corrected the measured position angles from our binary fitting routine using the detector orientation of $-0.430 \pm 0.076^{\circ}$ East of North derived from images of the  $\Theta^1$ Ori C field in \citet{MAIRE2015}. 
Each reduced, combined image was also corrected for the 
median parallactic angle during the image sequence to align with sky coordinates. 
We adopt position angle errors for the final images that reflect the full range of parallactic 
angles during each image sequence (a maximum of 1.8$^{\circ}$ for the $H$ and $K_s$ 
observations, see Table~\ref{astrometry_table}).

We do not use the $K_p$-band photometry from the archival Keck/NIRC2 images due to the use of 
the translucent coronagraphic mask in those observations. We measured the angular separation and
position angle of the secondary in each of the eight reduced images from their DS9 WCS coordinates. The mean 
and standard deviation of each parameter was calculated and each are provided in Table~\ref{astrometry_table}. 

\section{Orbit Analysis}\label{orbit_fit}

To determine the orbit of NO UMa, we follow standard binary orbit formalism (see Appendix~\ref{fit_formalism})
using the methods presented in \citet{ESPOSITO13}. 
In short, initial guesses of $P$, $T_0$, $e$, $K_1$, $K_2$, and $\gamma$ are made that are compatible with 
the observed data and correlations between orbital parameters. Then, a simultaneous astrometric and spectroscopic
best-fit orbit solution is solved using Levenberg-Marquardt least-squares minimization. We tested ranges of initial guesses to
investigate the effect on the resulting fit parameters. Due to the good coverage of both our astrometric and spectroscopic 
data over different phases of the orbit, we found that choices of initial guess comparable to previous estimates  
have no significant effect on the results of the fits.
The measured $t$, $x$, and $y$ inputs to the orbit analysis include those from our three 
LBTI/LMIRcam images, the Keck/NIRC2 image, and the six speckle interferometry measurements 
from \citet{BALEGA13}\footnote{The position angles of five of the six speckle interferometry points were 
rotated by 180$^{\circ}$ to converge on an orbital solution (see Table~\ref{astrometry_table}).}. 
The measured astrometry is provided in Table~\ref{astrometry_table}. The measured $v_1$ and $v_2$ values used in the fit are the 
STELLA-I/SES observations from \citet{STRASSMEIER12} described in \S~\ref{sec_props}. When exploring this RV data, we found that in a spectroscopic only fit, 
the measured RV errors resulted in correlated fit residuals for the primary and secondary that had standard deviations of 
$\sim$0.1 km s$^{-1}$ and $\sim$0.2 km s$^{-1}$, respectively. 
These SES systematics were previously investigated by \citet[][see their Fig.~1]{WEBER11} and are likely instrumental or calibration effects. 
To compensate for this underestimation of the true uncertainty, we added in quadrature the standard deviation
of the residuals from each of the fits back into the measured errors of each component as an additional error term.

\begin{figure}[!b]
\epsscale{1.0}
\plotone{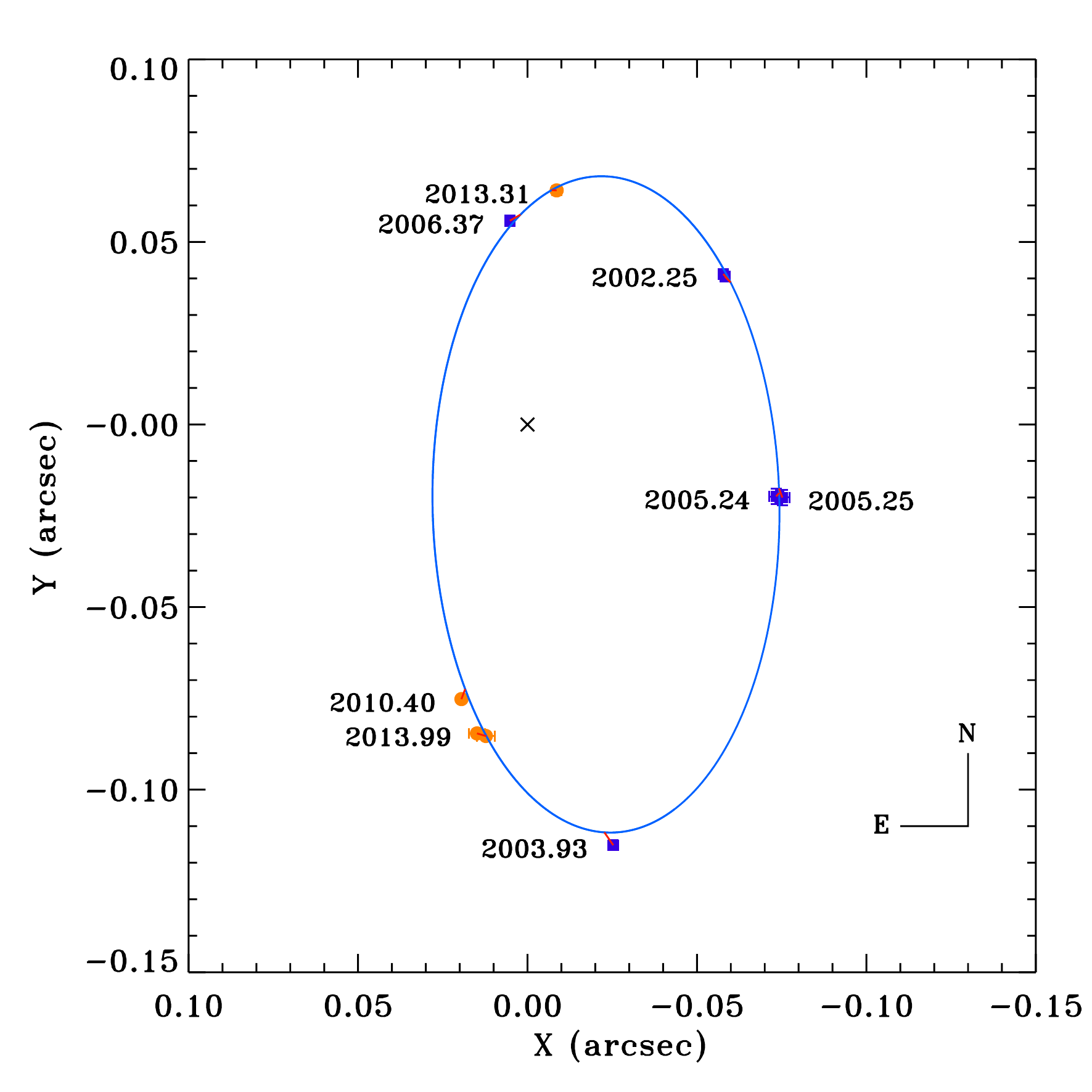}
\caption{Best-fit orbit of NO UMa B (blue ellipse) with respect to NO UMa A (cross).
The fit is compared to the relative astrometry from AO imaging (orange circles) and speckle interferometry 
(blue squares) measurements. Short red lines connect the data points to their expected positions along the orbit on the dates
of observation. The X and Y coordinates correspond to RA and Dec, respectively. The orbit is counter clockwise with a
period of $\sim$3.5 years. The error bars of some of the 
measurements are within the plot symbols. \label{orbit_image}}
\end{figure}

We present our best-fit orbital elements in the first column 
of Table~\ref{orbit_table} in comparison to the previous best estimates. The 1$\sigma$ 
uncertainties in the elements were estimated using MC methods where we drew $10^3$ random trials 
of the astrometric measurements from Gaussian error distributions around the nominal values and repeated the 
minimization procedure. Our independent estimates of the orbital parameters are  
consistent within 3$\sigma$ with those previously reported using both astrometric and spectroscopic data. Our constraints match
well with those derived by \citet{ARENOU00} and \citet{STRASSMEIER12}. Our more precise astrometric measurements 
allow us to place much tighter constraints on each of the orbital elements when compared to the \citet{ARENOU00} 
solution. Figure~\ref{orbit_image} shows our best-fit orbit compared to the observed astrometric data.

\begin{table*}
\begin{center}
\caption{Summary of NO UMa orbital properties  \label{tab1}}
\begin{tabular}{lccccc}
\hline\hline
 & This & Arenou et al. & Strassmeier et al. & Halbwachs et al. & Goldin \& Makarov \\
& Work & (2000) & (2012) & (2003) & (2007) \\
\hline
Period $P$ (days) & $1278.17\pm0.39$   & $1284.37\pm2.25$  & $1274.70\pm0.86$  & 1284.4 & $1366^{+199}_{-127}$ \\ 
Date of Periastron $T_0$ (days) & $2456490.9\pm0.3^b$  &  $2447512.91\pm6.50$$^a$  & $2455213.63\pm0.2$$^b$  & \dots & $171^{+564}_{-104}$$^c$ \\
Semi-major axis $a$ (mas)  & $101.4\pm0.1$ 	   &  $46.9\pm2.2$  & $39.3\pm0.9$/sin$i$   & \dots & $35.2^{+4.8}_{-3.2}$\\
Eccentricity $e$ & $0.508\pm0.001$ 	   &  $0.507\pm0.015$  & $0.5071\pm0.0006$  & 0.50 & $0.63^{+0.09}_{-0.07}$ \\
Inclination $i$ ($^{\circ}$) & $58.7\pm0.4$  & $60.1\pm3.5$  & \dots & \dots & $61\pm5$ \\
Argument of periastron $\omega$ ($^{\circ}$)  & $246.9\pm0.1$  & $248.8\pm2.5$  & $247.36\pm0.13$  & \dots & $63^{+12}_{-11}$ \\
PA of ascending node $\Omega$ ($^{\circ}$)   &  $356.0\pm0.3$  & $357.2\pm2.9$ & \dots  & \dots & $175^{+5}_{-6}$ \\
Primary semi-amp. $K_1$ (km s$^{-1}$)   &  $9.693 \pm 0.016$ &   \dots   &  $9.582 \pm 0.014$  &  9.78   &  \dots \\ 
Secondary semi-amp. $K_2$ (km s$^{-1}$)  & $12.447 \pm 0.023$ &   \dots   &  $12.470 \pm 0.021$  &  \dots   &  \dots \\ 
Systemic velocity $\gamma$ (km s$^{-1}$)  & $-9.873 \pm 0.007 $ &   \dots   &  $-9.862 \pm 0.004$  &  \dots   &  \dots \\ 
 \hline
System mass $M_{tot}$ ($M_{\odot}$) &  $1.47\pm0.03$  & $1.25\pm0.20$ & $0.9071\pm0.0025$/sin$^3$$i$  & 1.4  & \dots \\
Primary mass $M_{1}$ ($M_{\odot}$) & $0.83\pm0.02$ & $0.67\pm0.17$ & $0.5130\pm0.0021$/sin$^3$$i$&  0.75 & \dots \\
Secondary mass $M_{2}$ ($M_{\odot}$) & $0.64\pm0.02$ & $0.58\pm0.10$ & $0.3941\pm0.0014$/sin$^3$$i$  & 0.65  & \dots \\
Mass ratio $q$ ($M_{2}/M_{1}$)  &  $0.779\pm0.003$  & $0.87\pm0.26$ & $0.768\pm0.004$  &  0.86 & \dots \\
\hline
\label{orbit_table}
\end{tabular}
\tablecomments{$^a$Julian Date; $^b$Heliocentric Julian Date; $^c$Days after phase~=~0.}
\end{center}
\end{table*}

We used our best-fit orbital elements to derive the NO UMa component masses following equations~(11) and (12).  
 Propagating the associated uncertainties, we calculate $M_{1} = 0.83 \pm 0.02~M_{\odot}$ and 
$M_2 = 0.64 \pm 0.02~M_{\odot}$, respectively. We then combined the masses to find $M_{tot} = 1.47 \pm 0.03~M_{\odot}$. Our 
primary and secondary mass estimates are consistent with those of previous studies but with much smaller relative 
errors of $\sim$2\% and $\sim$3\%, respectively. We combined the inclination from our fit with the RV only 
mass constraints from \citet{STRASSMEIER12} and found system and component masses
in agreement with our estimates at better than 1$\sigma$. We also derived a new distance to the system, 
d = 25.87 $\pm$ 0.02 pc, slightly farther than the \emph{Hipparcos} measurement of $25.10\pm0.67$ pc 
\citep{VANLEEUWEN07}, but consistent within $\sim$1.2$\sigma$.

\begin{table}[!htb]
\caption{Astrometry for NO UMa}
\label{Table:astrometry}
\centering
\begin{tabular}{ccccc}
\hline
UT Date    	& Band &   $\rho$ &   PA   & O-C$_{X, Y}$   \\
(DD/MM/YYYY)	&	&		(mas)    &  ($^{\circ}$)  & (mas)     \\
\hline
\multicolumn{5}{c}{6-m BTA/Speckle Interferometry} \\
\hline
03/04/2002    & $\lambda$545$^a$  &    $71\pm2$    &   $305.5\pm1.0^b$   & 2.1, 2.3 \\
03/04/2002    & $\lambda$750$^a$  &    $71\pm2$    &   $304.8\pm1.4^b$   & 1.6, 1.6 \\
04/12/2003    & $\lambda$800$^a$  &    $118\pm2$    &   $192.4\pm1.4^b$  & -2.6, -3.5\\
27/03/2005    & $\lambda$545$^a$  &    $76\pm3$    &   $255.0\pm1.7^b$    & 0.9, -1.0 \\
31/03/2005    & $\lambda$800$^a$  &    $78\pm3$    &   $254.5\pm2.1^b$    & -1.0, -2.5\\
17/05/2006    & $\lambda$545$^a$  &    $56\pm2$    &   $5.3\pm1.4$   & 3.1, -1.7 \\
\hline
\multicolumn{5}{c}{Keck/NIRC2} \\
\hline
27/05/2010    & $K_p$  &    $77.7\pm0.7$    &   $165.5\pm0.3$   & 1.1, -2.8  \\
\hline  
\multicolumn{5}{c}{LBT/LMIRcam} \\ 
\hline
22/04/2013  & $L^{\prime}$    &   $64.7\pm0.2$    &   $352.4\pm1.3$ & -1.4, -0.1 \\
26/12/2013 &  $H$   &   $85.9\pm1.7$   &  $170.0\pm1.8$  & 2.8, 0.7 \\
26/12/2013    & $K_s$  &    $86.2\pm0.4$    &   $171.8\pm1.8$   & 0.2, 0.1  \\
\hline
\end{tabular}
\label{astrometry_table}
\tablecomments{$^a$Central wavelength in nm of the filter used during the 
observation. $^b$PA measurement from \citet{BALEGA13} rotated by 
180$^{\circ}$ to be consistent with previously estimated
orbit parameters.}
\end{table}

\section{Discussion}

\subsection{Physical Properties of NO UMa A and B}

To estimate the physical properties of NO UMa A and B, we use our measured values of 
$K_s$ and distance and $M_*$ and interpolate within the main-sequence color-temperature conversion 
table of \citet{PECAUT13}. To estimate the errors on our interpolated 
values, we employ MC techniques assuming Gaussian error 
distributions. The final values and their errors are the medians and dispersions of the interpolated distributions. 

We combine the measured $K_s$ magnitude of NO UMa A and the distance to the system 
to find an absolute $K_s$-band magnitude $\mathrm{M_{K_s} = 4.06\pm0.03}$. We use this 
measurement to estimate a SpTy and effective temperature of K2.0V$\pm$0.5 and 
$\mathrm{T_{eff} = 5010\pm50}$ K, respectively.  For NO UMa B, we estimate a SpTy of 
$\mathrm{K6.5V\pm0.5}$ and $\mathrm{T_{eff} = 4140\pm30}$ K from $\mathrm{M_{K_s} = 4.77\pm0.04}$. 
Using the measured masses of the A and B components, we estimate parameters in the same way and find values 
that are consistent with those estimated from $M_{K_s}$ but with larger uncertainties. The SpTy and effective temperature
estimates using $M_{K_s}$ and the masses are also consistent with the integrated 
SpTy of the system. Using the more precise $M_{K_s}$ temperature estimates, our $\mathrm{T_{eff}}$ of NO UMa A is consistent with the previous estimate from 
\citet{STRASSMEIER12} but our $\mathrm{T_{eff}}$ for the B component is $\sim$750 K cooler. 
This can likely be attributed to their use of a blended spectrum to estimate the temperature of each 
component rather than resolved measurements.

We also used our measured $M_{K_s}$ values to estimate component luminosities. Interpolating from the 
\citet{PECAUT13} table, we estimate $\mathrm{log(L/L_{\odot}) = -0.49\pm0.03}$ for NO UMa A and  
$\mathrm{log(L/L_{\odot}) = -0.97\pm0.02}$ for NO UMa B. These luminosities and the other
 physical parameters of the components estimated using $M_{K_s}$ are compiled in Table~\ref{properties_table}.

\subsection{NO UMa and the Ursa Major Moving Group}\label{group_membership}

References to a group of stars sharing similar kinematics in and around the constellation Ursa Major 
date back nearly 150 years \citep[e.g.][]{PROCTOR69}. A full history of the associated literature is 
beyond the scope of this paper but we do summarize some of the modern studies of the Ursa Major 
moving group (UMaG) and evaluate NO UMa's membership in it using our new data. 

A detailed, early study of the UMaG was presented by \citet{ROMAN49} who investigated the 
kinematics of all proposed members at that time. We point the interested reader to her exhaustive 
summary of previous works related to the group.  Roman's analysis revealed a compact nucleus of 
14 stars surrounded by a larger stream extending to radii of $\sim$100 pc. This early list already 
included NO UMa (listed as HD 109011) as a nuclear member of the group. The work of \citet{SODERBLOM93} 
used chromospheric activity (as traced by Ca II H \& K) and improved kinematic measurements to 
identify a list of 43 UMaG members. 

The most recent comprehensive studies of the group include those of \citet{KING03, KING05} 
and \citet{AMMLER09}.  \citet{KING03} sought to reinvestigate previously proposed members with new 
astrometric, photometric, and spectroscopic data.  From an input list of $\sim$220 proposed UMaG 
candidates, they identify 57 probable and possible members that are well defined in kinematic and 
color-magnitude space. Their results confirmed NO UMa as a bona-fide member of the UMaG nucleus.  
Comparison of evolution models to the empirical color-magnitude-diagram (CMD) of their refined membership lists suggested an 
age of 500$\pm$100 Myr for the group. The follow-up paper of \citet{KING05} examined activity, Li depletion, 
and abundances in members and affirms that the UMaG has approximately solar metallicity 
and its age overlaps with the Hyades and Coma Ber 
clusters within measurement and model uncertainties. \citet{AMMLER09} present a similar study and 
reach similar conclusions. \citet{TABERNERO14} perform a detailed chemical tagging 
study of candidate FGK type members of the UMaG and find 29 stars with similar chemical
compositions and [Fe/H] = $0.03 \pm 0.07$ dex. Using a Bayesian framework, \citet{BRANDT15} compared reliable, main-sequence-turn-off,
members of the UMaG to modern evolution models that include the effects of stellar rotation and affirmed the 
$\sim$500 Myr age of the group. Additional UMaG members are still being proposed, such as the candidate 
M dwarf members with measured parallaxes in \citet{SHKOLNIK12, RIEDEL14} and \citet{BOWLER15}. 

\citet{MAMAJEK10} present a comprehensive analysis of the UMaG's nuclear membership and kinematics  
using updated parallaxes from the \emph{Hipparcos} re-reduction of \citet{VANLEEUWEN07}. Here we 
re-evaluate the kinematics of NO UMa using the photocentric motion corrected proper motions from 
\cite{ARENOU00} and the new  systemic RV and distance from our orbit  fit.
 We calculate Cartesian Galactic velocities and positions, 
\emph{UVWXYZ}, using the methods outlined by \citet{JOHNSON87} updated for J2000.0 
coordinates\footnote{We define $U$ and $X$ positive toward the Galactic center, 
$V$ and $Y$ positive in the direction 
of solar motion around the galaxy, and $W$ and $Z$ positive toward the north Galactic pole. In this 
coordinate system, the Sun lies at the origin.}. We calculate 
$UVW_{NO~UMa}$ =  (14.19, 3.08, -7.70) $\pm$ (0.14, 0.07, 0.14) km s$^{-1}$ and 
$XYZ_{NO~UMa}$ = (-7.70, 9.51, 22.79) $\pm$ (0.01, 0.01, 0.01) pc. Our updated calculation of 
$UVW_{NO~UMa}$ is a close match to the UMaG nucleus mean velocity of 
$UVW_{UMaG}$ = (15.0, 2.8, -8.1) $\pm$ (0.4, 0.7, 1.0) km s$^{-1}$ \citep{MAMAJEK10}. 
When compared to the velocities calculated using the \emph{Hipparcos} proper motions and parallax
\citep{VANLEEUWEN07} and the mean system 
RV from \cite{GONTCHAROV06}, $UVW_{Hip}$ =  (15.9, -1.2, -9.9) $\pm$ (0.2, 0.3, 0.4) km s$^{-1}$, 
the revised velocities are much improved.

\begin{figure*}[!htb]
\epsscale{1.1}
\plotone{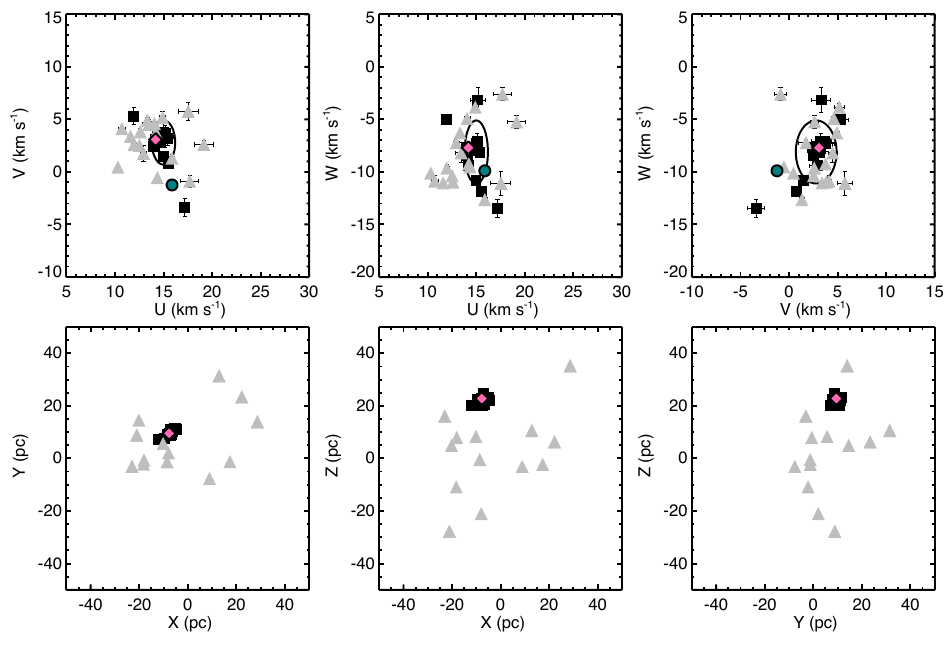}
\caption{Projections of the six-dimensional Galactic kinematics of the Ursa Major moving group (UMaG) compared to previous (teal circle) and revised (pink diamond) 
kinematics of NO UMa. Proposed nuclear (black squares) and stream (gray triangles) members are taken from the list of \citet{KING03}. (Top) Projections in 
$UVW$ Galactic velocity. The black ellipse designates the 3$\sigma$ dispersion of the average UMaG nucleus velocities from \citet{MAMAJEK10}. 
Our revised NO UMa $UVW$ velocities place it firmly in the UMaG nucleus. (Bottom) Projections of the $XYZ$ Galactic positions. NO UMa and other UMaG nucleus 
members occupy a very small volume of space, only $\sim$10 pc$^{3}$, and are surrounded by more distant kinematic stream members.  \label{UVW_fig}}
\end{figure*}

Figure~\ref{UVW_fig} shows projections of the six-dimensional 
Galactic kinematics of proposed nuclear and stream UMaG members 
from \cite{KING03} compared to our new estimates of NO UMa's kinematics and those calculated using 
previously available data. The new kinematic estimates place NO UMa firmly in the tight $UVW$ distribution of
the UMaG nucleus where previously available data placed it as a $\gtrsim$3$\sigma$ outlier in $U$, $V$, and $W$. 
\citet{MAMAJEK10} found that all the kinematic outliers in the UMaG nucleus were proposed 
binaries and suggested that their larger peculiar velocities may be attributed to binary motion. These results indicate 
that this hypothesis was true at least in the case of NO UMa. Our revised kinematics and the numerous age indicators
described in \S~\ref{sec_props} reinforce NO UMa's status as a bona-fide member of the UMaG nucleus.

\subsection{Model Comparisons}

Both components of the NO UMa system are expected to have settled onto the main-sequence at the 
500 Myr age of the UMa group and large deviations from model predictions are not expected. However, 
comparison of our measured masses to model estimates can provide insight into the scatter between the 
models and reveal any discrepancies between the measurements 
and model predictions. To explore these possibilities, we use our luminosities to estimate 
component masses from five sets of stellar evolution models at an age of 500 Myr. From the models of \citet[][Padova models]{BERTELLI08}, 
\citet[][Dartmouth models]{DOTTER08},  \citet[][PARSEC models]{BRESSAN12}, \citet[][Geneva models]{EKSTROEM12}, and \citet[][BHAC models]{BARAFFE15} 
with approximately solar abundances, we interpolate the masses and their 
uncertainties using MC methods and compare directly to our constraints. The model 
derived masses and model abundances are provided in Table~\ref{mass_table}.

\begin{table}
\begin{center}
\caption{Model Component Mass Estimates}
\begin{tabular}{lcc}
\hline\hline
  Model & Primary Mass & Secondary Mass  \\
            &  ($M_{\odot}$) &  ($M_{\odot}$)  \\
\hline
\citet{BERTELLI08} & $0.83\pm0.01$   & $0.65\pm0.01$  \\ 
($Z = 0.017, Y = 0.27$) &  &  \\
& & \\
\citet{DOTTER08}  & $0.85\pm0.01$  &  $0.66\pm0.01$ \\
($Z = 0.019, Y = 0.27$) &  &   \\
& & \\
\citet{BRESSAN12}  & $0.83\pm0.01$  & $0.66\pm0.01$   \\
($Z = 0.017, Y = 0.28$)  &   &  \\
& & \\
\citet{EKSTROEM12}  &  $0.84\pm0.01$  & $0.64\pm0.01$  \\
($Z = 0.014, Y = 0.27$) &  &  \\
& & \\
\citet{BARAFFE15}  &  $0.85\pm0.01$  & $0.67\pm0.01$  \\
($Z = 0.015, Y = 0.27$) &  &  \\
\hline
\label{mass_table}
\end{tabular}
\tablecomments{$Z$ = metallicity, $Y$ = He fraction}
\end{center}
\end{table}

\begin{figure}[!htb]
\epsscale{1.0}
\plotone{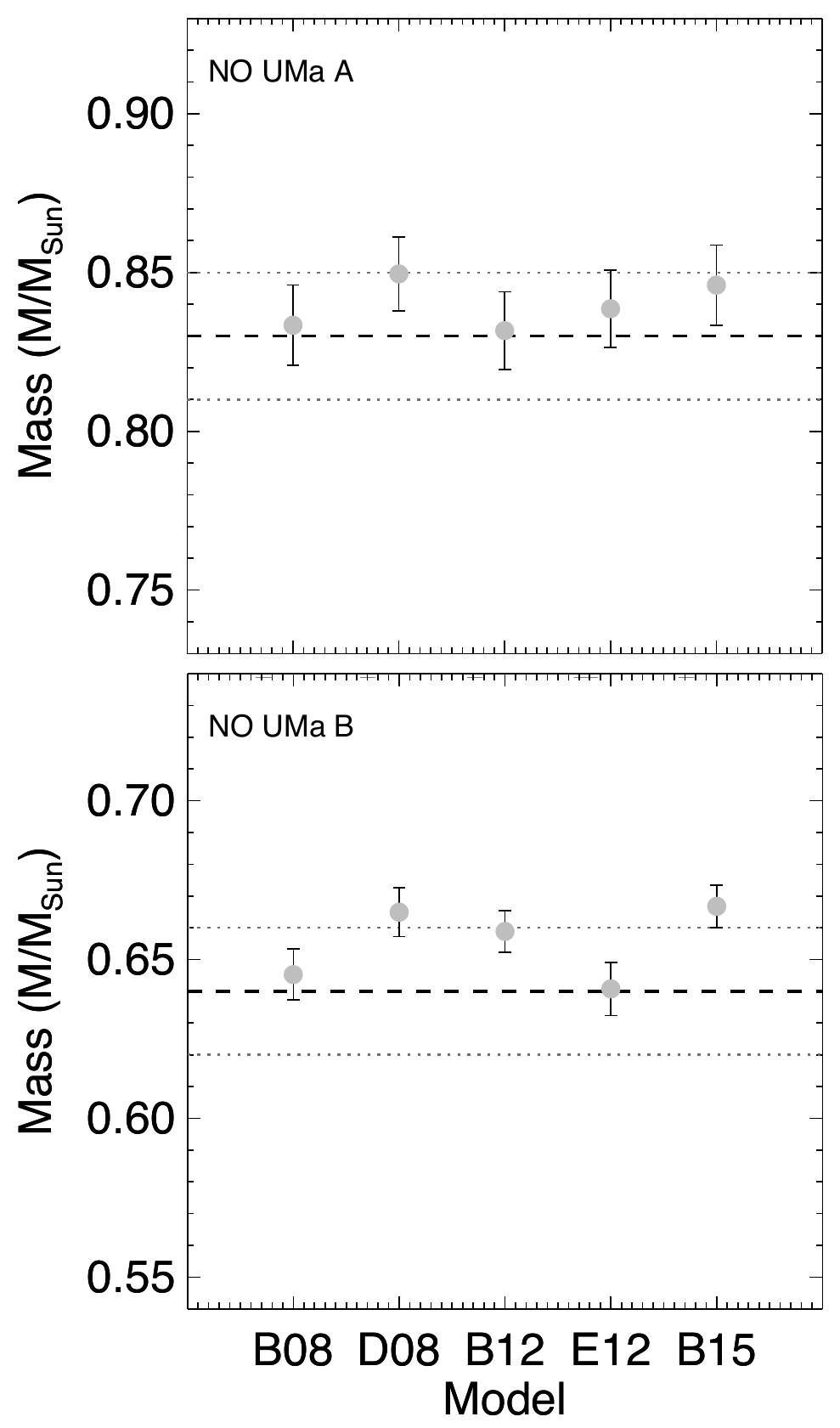}
\caption{Measured masses (dashed) and uncertainties (dotted) of NO UMa A (top) and B (bottom) compared to 
estimated masses from five sets of stellar evolution models (filled circles) at an age of 500 Myr: \citet[][B08]{BERTELLI08}, 
\citet[][D08]{DOTTER08},  \citet[][B12]{BRESSAN12}, \cite[][E12]{EKSTROEM12}, and \cite[][B15]{BARAFFE15}. The model estimated masses
match our orbit derived masses well within the uncertainties and there is little scatter between the models.  \label{mass_fig}}
\end{figure}

Figure~\ref{mass_fig} shows our measured masses compared to the predicted masses from the five 
sets of models. Each of the model derived masses match our measurements within the 1$\sigma$
uncertainties for each component. There is very little scatter between the masses estimated
from each set of models. The small observed scatter can be attributed to differences in abundances
and input physics. These comparisons indicate that modern stellar evolution models reproduce well the 
measured masses of intermediate age K-type stars.

\section{Conclusions}
   
Multi-epoch, infrared, AO imaging with LBTI/LMIRcam during the LEECH exoplanet imaging survey resolved the 
components of the known spectroscopic binary NO UMa at separations $<0\farcs09$. The binary exhibited about 180$^{\circ}$
of orbital motion over the $\sim$7 months spanning our images. We combined astrometry from our data with 
archival AO observations, published speckle interferometry measurements, and published velocity curves and performed
a simultaneous astrometric/spectroscopic orbit fit using 
minimization techniques. The complete set of estimated orbital parameters from our fit are consistent with previous
determinations that used data covering a smaller portion of the orbit and lead to significant improvements in precision. We 
estimated component masses of $\mathrm{0.83\pm0.02~M_{\odot}}$ and  $\mathrm{0.64\pm0.02~M_{\odot}}$ for the primary and secondary, 
respectively.  Our resolved, near-IR photometry, combined with a new system distance and empirical relations,
revealed a $\mathrm{K2.0V\pm0.5}$ primary and $\mathrm{K6.5V\pm0.5}$ secondary. The distance and systemic velocity from our orbit fit was combined
with binary motion corrected proper motions from the literature to revise the Galactic kinematics of NO UMa. We find that our kinematic estimates, that 
take binarity into account, are much improved over previous estimates that do not. This result suggests additional proposed binary UMaG members with large 
peculiar velocities can be reconciled if their kinematics are corrected for binary motion and strengthens NO UMa's status as a nuclear member of the 
$500 \pm 100$ Myr old UMaG. We compared our measured component masses to model estimates
from five sets of modern stellar evolution models at the nominal age of the system and found excellent agreement between 
the measured and model masses with little scatter between the models. NO UMa joins the short list of 
bright, nearby, short period binaries with known ages and fully characterized orbits. It is thus a late-type, mass and age benchmark. In addition 
to NO UMa, several other similar binaries have been resolved during LEECH observations and astrometric monitoring continues. 
 
\acknowledgments

We thank the anonymous referee for their constructive review that improved the 
quality of this manuscript. We thank the LBTI/LMIRcam instrument team for providing support during LEECH 
observations. A portion of the research of J.E.S was supported by an appointment to 
the NASA Postdoctoral Program at NASA Ames Research Center, administered by Oak 
Ridge Associated Universities through a contract with NASA. Support for A.J.S. was provided 
by the National Aeronautics and Space Administration through Hubble Fellowship grant 
HST-HF2-51349 awarded by the Space Telescope Science Institute, which is operated by 
the Association of Universities for Research in Astronomy, Inc., for NASA, under contract NAS 5-26555.  
 A.-L.M. and S.D. acknowledge support from the ``Progetti Premiali" funding scheme of the Italian 
 Ministry of Education, University, and Research. E.B. is supported by the Swiss National Science Foundation (SNSF). LEECH is funded by the NASA Origins of Solar 
Systems Program, grant NNX13AJ17G. The Large Binocular Telescope Interferometer is funded by NASA as part of its Exoplanet Exploration 
program. LMIRcam is funded by the National Science Foundation through grant NSF AST-0705296. STELLA was made possible by funding
through the State of Brandenburg (MWFK) and the German Federal Ministry of Education and Research (BMBF). The facility is a collaboration
of the AIP in Brandenburg with the IAC in Tenerife. This research has made use  of the SIMBAD database, operated at CDS, Strasbourg, France. This publication makes use 
of data products from the Wide-field Infrared Survey Explorer, which is a joint project of the 
University of California, Los Angeles, and the Jet Propulsion Laboratory/California Institute 
of Technology, funded by the National Aeronautics and Space Administration.  This publication makes use 
of data products from the Two Micron All Sky Survey, which is a joint project of the University of Massachusetts 
and the Infrared Processing and Analysis Center/California Institute of Technology, funded by the National 
Aeronautics and Space Administration and the National Science Foundation. This research has made use of the 
Keck Observatory Archive (KOA), which is operated by the W. M. Keck Observatory and the NASA Exoplanet 
Science Institute (NExScI), under contract with the National Aeronautics and Space Administration.



{\it Facilities:} \facility{LBT(LBTI/LMIRcam), STELLA-I(SES)}



\appendix

\section{Orbit Fit Formalism}\label{fit_formalism}

In an  astrometric (visual) binary system, the motion of the secondary relative to the primary is described by 
seven parameters called the Campbell elements: orbital period $P$, time of periastron passage 
$T_0$, semi-major axis $a$, eccentricity $e$, inclination $i$, argument of periastron $\omega$, 
and the position angle of the ascending node $\Omega$. If there exists a set of binary observations 
consisting of the time of observation $t$ and the relative coordinates of the secondary with respect 
to the primary $x$, $y$; and $P$, $T_0$, and $e$ are known, the remaining geometric elements of the 
orbit can be determined via minimization techniques through the Thiele-Innes elements: $A$, $B$, $F$, and $G$ 
\citep{HARTKOPF89, LUCY13, WOLLERT14}. This is possible 
because the Thiele-Innes elements are dependent on the orbital parameters:
\begin{eqnarray}
A &=& a \, (\cos{\omega}\cos{\Omega}-\sin{\omega}\sin{\Omega}\cos{i}) \\
B &=& a \, (\cos{\omega}\sin{\Omega}+\sin{\omega}\cos{\Omega}\cos{i}) \\
F &=& a \, (-\sin{\omega}\cos{\Omega}-\cos{\omega}\sin{\Omega}\cos{i}) \\
G &=& a \, (-\sin{\omega}\sin{\Omega}+\cos{\omega}\cos{\Omega}\cos{i}). 
\end{eqnarray} 
In practice, $P$, $T_0$, and $e$ are not typically known a priori. Thus, grid search techniques \citep{HARTKOPF89, SCHAEFER06}
or MC methods \citep{ESPOSITO13} are used to explore the parameter space and determine initial guesses 
that are compatible with the observations. Then, for every set of $P$, $T_0$, and $e$, the 
eccentric anomaly $E$ is given by Kepler's equation,
\begin{eqnarray}
E - e \sin{E} = \frac{2 \pi}{P}(t-T_0),
\end{eqnarray}
\noindent{and $x$ and $y$ at time $t$ are dependent on the Thiele-Innes elements, $E$, and $e$,} 
\begin{eqnarray}
x &=& A \cdot X + F \cdot Y \\
y &=& B \cdot X + G \cdot Y,
\end{eqnarray}
\noindent{where $X$~=~$\cos{E}-e$ and $Y$~=~$\sqrt{1-e^2} \cdot \sin{E}$}. Determination 
of the geometric orbit parameters from the Thiele-Innes elements is
then a problem of fitting a linear model to the observed data, minimizing the residuals between 
the model and observed $t$, $x$, and $y$, and solving the system of Thiele-Innes equations.

In a double-lined spectroscopic binary, velocity measurements of each component can independently yield
$P$, $T_0$, $e$, and $\omega$. The inclusion of these data also adds three additional orbital parameters;  
the velocity semi-amplitudes of the primary and secondary, $K_1$ and $K_2$, respectively, and the systemic 
velocity $\gamma$. The measured velocities of each component, $v_1$ and $v_2$,  
are given by

\begin{eqnarray}
v_1 = +K_1[e \cos{\omega} + \cos{(\nu + \omega)}] + \gamma \\
v_2 = -K_2[e \cos{\omega} + \cos{(\nu + \omega)}] + \gamma, 
\end{eqnarray}

\noindent{where the true anomaly $\nu$ can be calculated from the the orbital elements 
following,}

\begin{eqnarray}
\tan{\frac{\nu}{2}} = \sqrt{\frac{1+e}{1-e}} \tan{\frac{E}{2}}.
\end{eqnarray}

Thus, the addition of spectroscopic velocity data and a simultaneous astrometric/spectroscopic orbit fit allows 
determination of all 10 orbital elements
($P$, $T_0$, $e$, $a$, $i$, $\omega$, $\Omega$, $K_1$, $K_2$, $\gamma$) and 
a full orbit solution \citep{SCHAEFER08}. The resulting astrometric
and spectroscopic orbital parameters provide the masses of the primary and secondary

\begin{eqnarray}
M_1 = \frac{1.036 \times 10^{-7} (K_1 + K_2)^2 K_2 P (1-e^2)^{3/2}}{\sin^3i} \\
M_2 = \frac{1.036 \times 10^{-7} (K_1 + K_2)^2 K_1 P (1-e^2)^{3/2}}{\sin^3i},
\end{eqnarray}

\noindent{and the distance to the system}

\begin{eqnarray}
d = \frac{9.189 \times 10^{-5} (K_1 + K_2) P (1-e^2)^{1/2}}{a \sin i}, 
\end{eqnarray}

\noindent{where $K_1$ and $K_2$ are in km s$^{-1}$, $P$ is in days, and $a$ is in arcseconds and $M_1$ and $M_2$ are
in $M_{\odot}$ and $d$ is in pc \citep{SCHAEFER08}.}

\bibliography{NO_UMa_refs}

\begin{thebibliography}{}
\expandafter\ifx\csname natexlab\endcsname\relax\def\natexlab#1{#1}\fi

\bibitem[{{Allende Prieto}(2004)}]{ALLENDE04}
{Allende Prieto}, C. 2004, Astronomische Nachrichten, 325, 604

\bibitem[{{Ammler-von Eiff} \& {Guenther}(2009)}]{AMMLER09}
{Ammler-von Eiff}, M., \& {Guenther}, E.~W. 2009, \aap, 508, 677

\bibitem[{{Arenou} {et~al.}(2000){Arenou}, {Halbwachs}, {Mayor}, {Palasi}, \&
  {Udry}}]{ARENOU00}
{Arenou}, F., {Halbwachs}, J.-L., {Mayor}, M., {Palasi}, J., \& {Udry}, S.
  2000, in IAU Symposium, Vol. 200, IAU Symposium, 135P

\bibitem[{{Bailey} {et~al.}(2014){Bailey}, {Hinz}, {Puglisi}, {Esposito},
  {Vaitheeswaran}, {Skemer}, {Defr{\`e}re}, {Vaz}, \& {Leisenring}}]{BAILEY14}
{Bailey}, V.~P., {Hinz}, P.~M., {Puglisi}, A.~T., {et~al.} 2014, in Society of
  Photo-Optical Instrumentation Engineers (SPIE) Conference Series, Vol. 9148,
  Society of Photo-Optical Instrumentation Engineers (SPIE) Conference Series,
  3

\bibitem[{{Balega} {et~al.}(2013){Balega}, {Balega}, {Gasanova}, {Dyachenko},
  {Maksimov}, {Malogolovets}, {Rastegaev}, \& {Shkhagosheva}}]{BALEGA13}
{Balega}, I.~I., {Balega}, Y.~Y., {Gasanova}, L.~T., {et~al.} 2013,
  Astrophysical Bulletin, 68, 53

\bibitem[{{Baraffe} {et~al.}(1998){Baraffe}, {Chabrier}, {Allard}, \&
  {Hauschildt}}]{BARAFFE98}
{Baraffe}, I., {Chabrier}, G., {Allard}, F., \& {Hauschildt}, P.~H. 1998, \aap,
  337, 403

\bibitem[{{Baraffe} {et~al.}(2015){Baraffe}, {Homeier}, {Allard}, \&
  {Chabrier}}]{BARAFFE15}
{Baraffe}, I., {Homeier}, D., {Allard}, F., \& {Chabrier}, G. 2015, \aap, 577,
  A42

\bibitem[{{Baranne} {et~al.}(1979){Baranne}, {Mayor}, \& {Poncet}}]{BARANNE79}
{Baranne}, A., {Mayor}, M., \& {Poncet}, J.~L. 1979, Vistas in Astronomy, 23,
  279

\bibitem[{{Beckers}(1993)}]{BECKERS93}
{Beckers}, J.~M. 1993, \araa, 31, 13

\bibitem[{{Bertelli} {et~al.}(2008){Bertelli}, {Girardi}, {Marigo}, \&
  {Nasi}}]{BERTELLI08}
{Bertelli}, G., {Girardi}, L., {Marigo}, P., \& {Nasi}, E. 2008, \aap, 484, 815

\bibitem[{{Beuzit} {et~al.}(2008){Beuzit}, {Feldt}, {Dohlen}, {Mouillet},
  {Puget}, {Wildi}, {Abe}, {Antichi}, {Baruffolo}, {Baudoz}, {Boccaletti},
  {Carbillet}, {Charton}, {Claudi}, {Downing}, {Fabron}, {Feautrier},
  {Fedrigo}, {Fusco}, {Gach}, {Gratton}, {Henning}, {Hubin}, {Joos}, {Kasper},
  {Langlois}, {Lenzen}, {Moutou}, {Pavlov}, {Petit}, {Pragt}, {Rabou}, {Rigal},
  {Roelfsema}, {Rousset}, {Saisse}, {Schmid}, {Stadler}, {Thalmann}, {Turatto},
  {Udry}, {Vakili}, \& {Waters}}]{BEUZIT08}
{Beuzit}, J.-L., {Feldt}, M., {Dohlen}, K., {et~al.} 2008, in Society of
  Photo-Optical Instrumentation Engineers (SPIE) Conference Series, Vol. 7014,
  Society of Photo-Optical Instrumentation Engineers (SPIE) Conference Series,
  18

\bibitem[{{Bonnefoy} {et~al.}(2014){Bonnefoy}, {Currie}, {Marleau},
  {Schlieder}, {Wisniewski}, {Carson}, {Covey}, {Henning}, {Biller}, {Hinz},
  {Klahr}, {Marsh Boyer}, {Zimmerman}, {Janson}, {McElwain}, {Mordasini},
  {Skemer}, {Bailey}, {Defr{\`e}re}, {Thalmann}, {Skrutskie}, {Allard},
  {Homeier}, {Tamura}, {Feldt}, {Cumming}, {Grady}, {Brandner}, {Helling},
  {Witte}, {Hauschildt}, {Kandori}, {Kuzuhara}, {Fukagawa}, {Kwon}, {Kudo},
  {Hashimoto}, {Kusakabe}, {Abe}, {Brandt}, {Egner}, {Guyon}, {Hayano},
  {Hayashi}, {Hayashi}, {Hodapp}, {Ishii}, {Iye}, {Knapp}, {Matsuo}, {Mede},
  {Miyama}, {Morino}, {Moro-Martin}, {Nishimura}, {Pyo}, {Serabyn}, {Suenaga},
  {Suto}, {Suzuki}, {Takahashi}, {Takami}, {Takato}, {Terada}, {Tomono},
  {Turner}, {Watanabe}, {Yamada}, {Takami}, \& {Usuda}}]{BONNEFOY14}
{Bonnefoy}, M., {Currie}, T., {Marleau}, G.-D., {et~al.} 2014, \aap, 562, A111

\bibitem[{{Bowler} {et~al.}(2015){Bowler}, {Liu}, {Shkolnik}, \&
  {Tamura}}]{BOWLER15}
{Bowler}, B.~P., {Liu}, M.~C., {Shkolnik}, E.~L., \& {Tamura}, M. 2015, \apjs,
  216, 7

\bibitem[{{Brandt} \& {Huang}(2015)}]{BRANDT15}
{Brandt}, T.~D., \& {Huang}, C.~X. 2015, ArXiv e-prints, arXiv:1501.04404

\bibitem[{{Bressan} {et~al.}(2012){Bressan}, {Marigo}, {Girardi}, {Salasnich},
  {Dal Cero}, {Rubele}, \& {Nanni}}]{BRESSAN12}
{Bressan}, A., {Marigo}, P., {Girardi}, L., {et~al.} 2012, \mnras, 427, 127

\bibitem[{{Burrows} {et~al.}(1997){Burrows}, {Marley}, {Hubbard}, {Lunine},
  {Guillot}, {Saumon}, {Freedman}, {Sudarsky}, \& {Sharp}}]{BURROWS97}
{Burrows}, A., {Marley}, M., {Hubbard}, W.~B., {et~al.} 1997, \apj, 491, 856

\bibitem[{{Cutri} \& {et al.}(2013)}]{CUTRI13}
{Cutri}, R.~M., \& {et al.} 2013, VizieR Online Data Catalog, 2328, 0

\bibitem[{{Cutri} {et~al.}(2003){Cutri}, {Skrutskie}, {van Dyk}, {Beichman},
  {Carpenter}, {Chester}, {Cambresy}, {Evans}, {Fowler}, {Gizis}, {Howard},
  {Huchra}, {Jarrett}, {Kopan}, {Kirkpatrick}, {Light}, {Marsh}, {McCallon},
  {Schneider}, {Stiening}, {Sykes}, {Weinberg}, {Wheaton}, {Wheelock}, \&
  {Zacarias}}]{CUTRI03}
{Cutri}, R.~M., {Skrutskie}, M.~F., {van Dyk}, S., {et~al.} 2003, {2MASS All
  Sky Catalog of point sources.}

\bibitem[{{Delorme} {et~al.}(2013){Delorme}, {Gagn{\'e}}, {Girard}, {Lagrange},
  {Chauvin}, {Naud}, {Lafreni{\`e}re}, {Doyon}, {Riedel}, {Bonnefoy}, \&
  {Malo}}]{DELORME13}
{Delorme}, P., {Gagn{\'e}}, J., {Girard}, J.~H., {et~al.} 2013, \aap, 553, L5

\bibitem[{{Dotter} {et~al.}(2008){Dotter}, {Chaboyer}, {Jevremovi{\'c}},
  {Kostov}, {Baron}, \& {Ferguson}}]{DOTTER08}
{Dotter}, A., {Chaboyer}, B., {Jevremovi{\'c}}, D., {et~al.} 2008, \apjs, 178,
  89

\bibitem[{{Duch{\^e}ne} \& {Kraus}(2013)}]{DUCHENE13}
{Duch{\^e}ne}, G., \& {Kraus}, A. 2013, \araa, 51, 269

\bibitem[{{Duquennoy} {et~al.}(1991){Duquennoy}, {Mayor}, \&
  {Halbwachs}}]{DUQUENNOY91}
{Duquennoy}, A., {Mayor}, M., \& {Halbwachs}, J.-L. 1991, \aaps, 88, 281

\bibitem[{{Dutrey} {et~al.}(2014){Dutrey}, {di Folco}, {Guilloteau}, {Boehler},
  {Bary}, {Beck}, {Beust}, {Chapillon}, {Gueth}, {Hur{\'e}}, {Pierens},
  {Pi{\'e}tu}, {Simon}, \& {Tang}}]{DUTREY14}
{Dutrey}, A., {di Folco}, E., {Guilloteau}, S., {et~al.} 2014, \nat, 514, 600

\bibitem[{{Ekstr{\"o}m} {et~al.}(2012){Ekstr{\"o}m}, {Georgy}, {Eggenberger},
  {Meynet}, {Mowlavi}, {Wyttenbach}, {Granada}, {Decressin}, {Hirschi},
  {Frischknecht}, {Charbonnel}, \& {Maeder}}]{EKSTROEM12}
{Ekstr{\"o}m}, S., {Georgy}, C., {Eggenberger}, P., {et~al.} 2012, \aap, 537,
  A146

\bibitem[{{Esposito} {et~al.}(2010){Esposito}, {Riccardi},
  {Quir{\'o}s-Pacheco}, {Pinna}, {Puglisi}, {Xompero}, {Briguglio}, {Busoni},
  {Fini}, {Stefanini}, {Brusa}, {Tozzi}, {Ranfagni}, {Pieralli}, {Guerra},
  {Arcidiacono}, \& {Salinari}}]{ESPOSITO10}
{Esposito}, S., {Riccardi}, A., {Quir{\'o}s-Pacheco}, F., {et~al.} 2010, \ao,
  49, G174

\bibitem[{{Esposito} {et~al.}(2011){Esposito}, {Riccardi}, {Pinna}, {Puglisi},
  {Quir{\'o}s-Pacheco}, {Arcidiacono}, {Xompero}, {Briguglio}, {Agapito},
  {Busoni}, {Fini}, {Argomedo}, {Gherardi}, {Brusa}, {Miller}, {Guerra},
  {Stefanini}, \& {Salinari}}]{ESPOSITO11}
{Esposito}, S., {Riccardi}, A., {Pinna}, E., {et~al.} 2011, in Society of
  Photo-Optical Instrumentation Engineers (SPIE) Conference Series, Vol. 8149,
  Society of Photo-Optical Instrumentation Engineers (SPIE) Conference Series

\bibitem[{{Esposito} {et~al.}(2013){Esposito}, {Mesa}, {Skemer}, {Arcidiacono},
  {Claudi}, {Desidera}, {Gratton}, {Mannucci}, {Marzari}, {Masciadri}, {Close},
  {Hinz}, {Kulesa}, {McCarthy}, {Males}, {Agapito}, {Argomedo}, {Boutsia},
  {Briguglio}, {Brusa}, {Busoni}, {Cresci}, {Fini}, {Fontana}, {Guerra},
  {Hill}, {Miller}, {Paris}, {Pinna}, {Puglisi}, {Quiros-Pacheco}, {Riccardi},
  {Stefanini}, {Testa}, {Xompero}, \& {Woodward}}]{ESPOSITO13}
{Esposito}, S., {Mesa}, D., {Skemer}, A., {et~al.} 2013, \aap, 549, A52

\bibitem[{{Goldin} \& {Makarov}(2007)}]{GOLDIN07}
{Goldin}, A., \& {Makarov}, V.~V. 2007, \apjs, 173, 137

\bibitem[{{Gontcharov}(2006)}]{GONTCHAROV06}
{Gontcharov}, G.~A. 2006, Astronomy Letters, 32, 759

\bibitem[{{Halbwachs} {et~al.}(2000){Halbwachs}, {Arenou}, {Mayor}, \&
  {Udry}}]{HALBWACHS00}
{Halbwachs}, J.-., {Arenou}, F., {Mayor}, M., \& {Udry}, S. 2000, in IAU
  Symposium, Vol. 200, IAU Symposium, 132P

\bibitem[{{Halbwachs} {et~al.}(2003){Halbwachs}, {Mayor}, {Udry}, \&
  {Arenou}}]{HALBWACHS03}
{Halbwachs}, J.~L., {Mayor}, M., {Udry}, S., \& {Arenou}, F. 2003, \aap, 397,
  159

\bibitem[{{Hartkopf} {et~al.}(1989){Hartkopf}, {McAlister}, \&
  {Franz}}]{HARTKOPF89}
{Hartkopf}, W.~I., {McAlister}, H.~A., \& {Franz}, O.~G. 1989, \aj, 98, 1014

\bibitem[{{Heinze} \& {Hinz}(2005)}]{HEINZE05}
{Heinze}, A.~N., \& {Hinz}, P.~M. 2005, \aj, 130, 1929

\bibitem[{{Hinkley} {et~al.}(2011){Hinkley}, {Oppenheimer}, {Zimmerman},
  {Brenner}, {Parry}, {Crepp}, {Vasisht}, {Ligon}, {King}, {Soummer},
  {Sivaramakrishnan}, {Beichman}, {Shao}, {Roberts}, {Bouchez}, {Dekany},
  {Pueyo}, {Roberts}, {Lockhart}, {Zhai}, {Shelton}, \& {Burruss}}]{HINKLEY11}
{Hinkley}, S., {Oppenheimer}, B.~R., {Zimmerman}, N., {et~al.} 2011, \pasp,
  123, 74

\bibitem[{{Hinz} {et~al.}(2008){Hinz}, {Bippert-Plymate}, {Breuninger},
  {Connors}, {Duffy}, {Esposito}, {Hoffmann}, {Kim}, {Kraus}, {McMahon},
  {Montoya}, {Nash}, {Durney}, {Solheid}, {Tozzi}, \& {Vaitheeswaran}}]{HINZ08}
{Hinz}, P.~M., {Bippert-Plymate}, T., {Breuninger}, A., {et~al.} 2008, in
  Society of Photo-Optical Instrumentation Engineers (SPIE) Conference Series,
  Vol. 7013, Society of Photo-Optical Instrumentation Engineers (SPIE)
  Conference Series, 28

\bibitem[{{H{\o}g} {et~al.}(2000){H{\o}g}, {Fabricius}, {Makarov}, {Urban},
  {Corbin}, {Wycoff}, {Bastian}, {Schwekendiek}, \& {Wicenec}}]{HOG00}
{H{\o}g}, E., {Fabricius}, C., {Makarov}, V.~V., {et~al.} 2000, \aap, 355, L27

\bibitem[{{Johnson} \& {Soderblom}(1987)}]{JOHNSON87}
{Johnson}, D.~R.~H., \& {Soderblom}, D.~R. 1987, \aj, 93, 864

\bibitem[{{Johnson} \& {Morgan}(1953)}]{JOHNSON53}
{Johnson}, H.~L., \& {Morgan}, W.~W. 1953, \apj, 117, 313

\bibitem[{{Jovanovic} {et~al.}(2013){Jovanovic}, {Weber}, \& {Allende
  Prieto}}]{JOVANOVIC13}
{Jovanovic}, M., {Weber}, M., \& {Allende Prieto}, C. 2013, Publications de
  l'Observatoire Astronomique de Beograd, 92, 169

\bibitem[{{King} \& {Schuler}(2005)}]{KING05}
{King}, J.~R., \& {Schuler}, S.~C. 2005, \pasp, 117, 911

\bibitem[{{King} {et~al.}(2003){King}, {Villarreal}, {Soderblom}, {Gulliver},
  \& {Adelman}}]{KING03}
{King}, J.~R., {Villarreal}, A.~R., {Soderblom}, D.~R., {Gulliver}, A.~F., \&
  {Adelman}, S.~J. 2003, \aj, 125, 1980

\bibitem[{{Kley} \& {Haghighipour}(2014)}]{KLEY14}
{Kley}, W., \& {Haghighipour}, N. 2014, \aap, 564, A72

\bibitem[{{Kraus} {et~al.}(2014){Kraus}, {Ireland}, {Cieza}, {Hinkley},
  {Dupuy}, {Bowler}, \& {Liu}}]{KRAUS14}
{Kraus}, A.~L., {Ireland}, M.~J., {Cieza}, L.~A., {et~al.} 2014, \apj, 781, 20

\bibitem[{{Leisenring} {et~al.}(2012){Leisenring}, {Skrutskie}, {Hinz},
  {Skemer}, {Bailey}, {Eisner}, {Garnavich}, {Hoffmann}, {Jones}, {Kenworthy},
  {Kuzmenko}, {Meyer}, {Nelson}, {Rodigas}, {Wilson}, \&
  {Vaitheeswaran}}]{LEISENRING12}
{Leisenring}, J.~M., {Skrutskie}, M.~F., {Hinz}, P.~M., {et~al.} 2012, in
  Society of Photo-Optical Instrumentation Engineers (SPIE) Conference Series,
  Vol. 8446, Society of Photo-Optical Instrumentation Engineers (SPIE)
  Conference Series

\bibitem[{{Lucy}(2013)}]{LUCY13}
{Lucy}, L.~B. 2013, \aap, 551, A47

\bibitem[{{Macintosh} {et~al.}(2008){Macintosh}, {Graham}, {Palmer}, {Doyon},
  {Dunn}, {Gavel}, {Larkin}, {Oppenheimer}, {Saddlemyer}, {Sivaramakrishnan},
  {Wallace}, {Bauman}, {Erickson}, {Marois}, {Poyneer}, \&
  {Soummer}}]{MACINTOSH08}
{Macintosh}, B.~A., {Graham}, J.~R., {Palmer}, D.~W., {et~al.} 2008, in Society
  of Photo-Optical Instrumentation Engineers (SPIE) Conference Series, Vol.
  7015, Society of Photo-Optical Instrumentation Engineers (SPIE) Conference
  Series, 18

\bibitem[{{Maire} {et~al.}(2015){Maire}, {Skemer}, {Hinz}, {Desidera},
  {Esposito}, {Gratton}, {Marzari}, {Skrutskie}, {Biller}, {Defr{\`e}re},
  {Bailey}, {Leisenring}, {Apai}, {Bonnefoy}, {Brandner}, {Buenzli}, {Claudi},
  {Close}, {Crepp}, {De Rosa}, {Eisner}, {Fortney}, {Henning}, {Hofmann},
  {Kopytova}, {Males}, {Mesa}, {Morzinski}, {Oza}, {Patience}, {Pinna},
  {Rajan}, {Schertl}, {Schlieder}, {Su}, {Vaz}, {Ward-Duong}, {Weigelt}, \&
  {Woodward}}]{MAIRE2015}
{Maire}, A.-L., {Skemer}, A.~J., {Hinz}, P.~M., {et~al.} 2015, \aap, 576, A133

\bibitem[{{Mamajek} \& {Hillenbrand}(2008)}]{MAMAJEK08}
{Mamajek}, E.~E., \& {Hillenbrand}, L.~A. 2008, \apj, 687, 1264

\bibitem[{{Mamajek} {et~al.}(2010){Mamajek}, {Kenworthy}, {Hinz}, \&
  {Meyer}}]{MAMAJEK10}
{Mamajek}, E.~E., {Kenworthy}, M.~A., {Hinz}, P.~M., \& {Meyer}, M.~R. 2010,
  \aj, 139, 919

\bibitem[{{Marleau} \& {Cumming}(2014)}]{MARLEAU14}
{Marleau}, G.-D., \& {Cumming}, A. 2014, \mnras, 437, 1378

\bibitem[{{Mayor} {et~al.}(1992){Mayor}, {Duquennoy}, {Halbwachs}, \&
  {Mermilliod}}]{MAYOR1992}
{Mayor}, M., {Duquennoy}, A., {Halbwachs}, J.-L., \& {Mermilliod}, J.-C. 1992,
  in Astronomical Society of the Pacific Conference Series, Vol.~32, IAU
  Colloq. 135: Complementary Approaches to Double and Multiple Star Research,
  ed. H.~A. {McAlister} \& W.~I. {Hartkopf}, 73

\bibitem[{{Mermilliod} \& {Mermilliod}(1994)}]{MERMILLIOD94}
{Mermilliod}, J.-C., \& {Mermilliod}, M. 1994, {Catalogue of Mean UBV Data on
  Stars}

\bibitem[{{Pecaut} \& {Mamajek}(2013)}]{PECAUT13}
{Pecaut}, M.~J., \& {Mamajek}, E.~E. 2013, \apjs, 208, 9

\bibitem[{{Perryman} {et~al.}(1997){Perryman}, {Lindegren}, {Kovalevsky},
  {Hoeg}, {Bastian}, {Bernacca}, {Cr{\'e}z{\'e}}, {Donati}, {Grenon},
  {Grewing}, {van Leeuwen}, {van der Marel}, {Mignard}, {Murray}, {Le Poole},
  {Schrijver}, {Turon}, {Arenou}, {Froeschl{\'e}}, \& {Petersen}}]{PERRYMAN97}
{Perryman}, M.~A.~C., {Lindegren}, L., {Kovalevsky}, J., {et~al.} 1997, \aap,
  323, L49

\bibitem[{{Proctor}(1869)}]{PROCTOR69}
{Proctor}, R.~A. 1869, Royal Society of London Proceedings Series I, 18, 169

\bibitem[{{Raghavan} {et~al.}(2010){Raghavan}, {McAlister}, {Henry}, {Latham},
  {Marcy}, {Mason}, {Gies}, {White}, \& {ten Brummelaar}}]{RAGHAVAN10}
{Raghavan}, D., {McAlister}, H.~A., {Henry}, T.~J., {et~al.} 2010, \apjs, 190,
  1

\bibitem[{{Rapson} {et~al.}(2015){Rapson}, {Kastner}, {Andrews}, {Hines},
  {Macintosh}, {Millar-Blanchaer}, \& {Tamura}}]{RAPSON15}
{Rapson}, V.~A., {Kastner}, J.~H., {Andrews}, S.~M., {et~al.} 2015, \apjl, 803,
  L10

\bibitem[{{Riccardi} {et~al.}(2010){Riccardi}, {Xompero}, {Briguglio},
  {Quir{\'o}s-Pacheco}, {Busoni}, {Fini}, {Puglisi}, {Esposito}, {Arcidiacono},
  {Pinna}, {Ranfagni}, {Salinari}, {Brusa}, {Demers}, {Biasi}, \&
  {Gallieni}}]{RICCARDI10}
{Riccardi}, A., {Xompero}, M., {Briguglio}, R., {et~al.} 2010, in Society of
  Photo-Optical Instrumentation Engineers (SPIE) Conference Series, Vol. 7736,
  Society of Photo-Optical Instrumentation Engineers (SPIE) Conference Series,
  2

\bibitem[{{Riedel} {et~al.}(2014){Riedel}, {Finch}, {Henry}, {Subasavage},
  {Jao}, {Malo}, {Rodriguez}, {White}, {Gies}, {Dieterich}, {Winters},
  {Davison}, {Nelan}, {Blunt}, {Cruz}, {Rice}, \& {Ianna}}]{RIEDEL14}
{Riedel}, A.~R., {Finch}, C.~T., {Henry}, T.~J., {et~al.} 2014, \aj, 147, 85

\bibitem[{{Roman}(1949)}]{ROMAN49}
{Roman}, N.~G. 1949, \apj, 110, 205

\bibitem[{{Schaefer} {et~al.}(2006){Schaefer}, {Simon}, {Beck}, {Nelan}, \&
  {Prato}}]{SCHAEFER06}
{Schaefer}, G.~H., {Simon}, M., {Beck}, T.~L., {Nelan}, E., \& {Prato}, L.
  2006, \aj, 132, 2618

\bibitem[{{Schaefer} {et~al.}(2008){Schaefer}, {Simon}, {Prato}, \&
  {Barman}}]{SCHAEFER08}
{Schaefer}, G.~H., {Simon}, M., {Prato}, L., \& {Barman}, T. 2008, \aj, 135,
  1659

\bibitem[{{Schlieder} {et~al.}(2014){Schlieder}, {Bonnefoy}, {Herbst},
  {L{\'e}pine}, {Berger}, {Henning}, {Skemer}, {Chauvin}, {Rice}, {Biller},
  {Girard}, {Lagrange}, {Hinz}, {Defr{\`e}re}, {Bergfors}, {Brandner},
  {Lacour}, {Skrutskie}, \& {Leisenring}}]{SCHLIEDER14}
{Schlieder}, J.~E., {Bonnefoy}, M., {Herbst}, T.~M., {et~al.} 2014, \apj, 783,
  27

\bibitem[{{Shkolnik} {et~al.}(2012){Shkolnik}, {Anglada-Escud{\'e}}, {Liu},
  {Bowler}, {Weinberger}, {Boss}, {Reid}, \& {Tamura}}]{SHKOLNIK12}
{Shkolnik}, E.~L., {Anglada-Escud{\'e}}, G., {Liu}, M.~C., {et~al.} 2012, \apj,
  758, 56

\bibitem[{{Skemer} {et~al.}(2012){Skemer}, {Hinz}, {Esposito}, {Burrows},
  {Leisenring}, {Skrutskie}, {Desidera}, {Mesa}, {Arcidiacono}, {Mannucci},
  {Rodigas}, {Close}, {McCarthy}, {Kulesa}, {Agapito}, {Apai}, {Argomedo},
  {Bailey}, {Boutsia}, {Briguglio}, {Brusa}, {Busoni}, {Claudi}, {Eisner},
  {Fini}, {Follette}, {Garnavich}, {Gratton}, {Guerra}, {Hill}, {Hoffmann},
  {Jones}, {Krejny}, {Males}, {Masciadri}, {Meyer}, {Miller}, {Morzinski},
  {Nelson}, {Pinna}, {Puglisi}, {Quanz}, {Quiros-Pacheco}, {Riccardi},
  {Stefanini}, {Vaitheeswaran}, {Wilson}, \& {Xompero}}]{SKEMER12}
{Skemer}, A.~J., {Hinz}, P.~M., {Esposito}, S., {et~al.} 2012, \apj, 753, 14

\bibitem[{{Skemer} {et~al.}(2014{\natexlab{a}}){Skemer}, {Marley}, {Hinz},
  {Morzinski}, {Skrutskie}, {Leisenring}, {Close}, {Saumon}, {Bailey},
  {Briguglio}, {Defrere}, {Esposito}, {Follette}, {Hill}, {Males}, {Puglisi},
  {Rodigas}, \& {Xompero}}]{SKEMER14}
{Skemer}, A.~J., {Marley}, M.~S., {Hinz}, P.~M., {et~al.} 2014{\natexlab{a}},
  \apj, 792, 17

\bibitem[{{Skemer} {et~al.}(2014{\natexlab{b}}){Skemer}, {Hinz}, {Esposito},
  {Skrutskie}, {Defr{\`e}re}, {Bailey}, {Leisenring}, {Apai}, {Biller},
  {Bonnefoy}, {Brandner}, {Buenzli}, {Close}, {Crepp}, {De Rosa}, {Desidera},
  {Eisner}, {Fortney}, {Henning}, {Hofmann}, {Kopytova}, {Maire}, {Males},
  {Millan-Gabet}, {Morzinski}, {Oza}, {Patience}, {Rajan}, {Rieke}, {Schertl},
  {Schlieder}, {Su}, {Vaz}, {Ward-Duong}, {Weigelt}, {Woodward}, \&
  {Zimmerman}}]{SKEMER14SPIE}
{Skemer}, A.~J., {Hinz}, P., {Esposito}, S., {et~al.} 2014{\natexlab{b}}, in
  Society of Photo-Optical Instrumentation Engineers (SPIE) Conference Series,
  Vol. 9148, Society of Photo-Optical Instrumentation Engineers (SPIE)
  Conference Series, 0

\bibitem[{{Skrutskie} {et~al.}(2010){Skrutskie}, {Jones}, {Hinz}, {Garnavich},
  {Wilson}, {Nelson}, {Solheid}, {Durney}, {Hoffmann}, {Vaitheeswaran},
  {McMahon}, {Leisenring}, \& {Wong}}]{SKRUTSKIE10}
{Skrutskie}, M.~F., {Jones}, T., {Hinz}, P., {et~al.} 2010, in Society of
  Photo-Optical Instrumentation Engineers (SPIE) Conference Series, Vol. 7735,
  Society of Photo-Optical Instrumentation Engineers (SPIE) Conference Series

\bibitem[{{Soderblom} \& {Mayor}(1993)}]{SODERBLOM93}
{Soderblom}, D.~R., \& {Mayor}, M. 1993, \aj, 105, 226

\bibitem[{{Spiegel} \& {Burrows}(2012)}]{SPIEGEL12}
{Spiegel}, D.~S., \& {Burrows}, A. 2012, \apj, 745, 174

\bibitem[{{Strassmeier} {et~al.}(2000){Strassmeier}, {Washuettl}, {Granzer},
  {Scheck}, \& {Weber}}]{STRASSMEIER00}
{Strassmeier}, K., {Washuettl}, A., {Granzer}, T., {Scheck}, M., \& {Weber}, M.
  2000, \aaps, 142, 275

\bibitem[{{Strassmeier} {et~al.}(2012){Strassmeier}, {Weber}, {Granzer}, \&
  {J{\"a}rvinen}}]{STRASSMEIER12}
{Strassmeier}, K.~G., {Weber}, M., {Granzer}, T., \& {J{\"a}rvinen}, S. 2012,
  Astronomische Nachrichten, 333, 663

\bibitem[{{Tabernero} {et~al.}(2014){Tabernero}, {Montes}, {Gonzalez
  Hernandez}, \& {Ammler-von Eiff}}]{TABERNERO14}
{Tabernero}, H.~M., {Montes}, D., {Gonzalez Hernandez}, J.~I., \& {Ammler-von
  Eiff}, M. 2014, ArXiv e-prints, arXiv:1409.2348

\bibitem[{{Tang} {et~al.}(2014){Tang}, {Dutrey}, {Guilloteau}, {Pi{\'e}tu}, {Di
  Folco}, {Beck}, {Ho}, {Boehler}, {Gueth}, {Bary}, \& {Simon}}]{TANG14}
{Tang}, Y.-W., {Dutrey}, A., {Guilloteau}, S., {et~al.} 2014, \apj, 793, 10

\bibitem[{{Thalmann} {et~al.}(2014){Thalmann}, {Desidera}, {Bonavita},
  {Janson}, {Usuda}, {Henning}, {K{\"o}hler}, {Carson}, {Boccaletti},
  {Bergfors}, {Brandner}, {Feldt}, {Goto}, {Klahr}, {Marzari}, \&
  {Mordasini}}]{THALMANN14}
{Thalmann}, C., {Desidera}, S., {Bonavita}, M., {et~al.} 2014, \aap, 572, A91

\bibitem[{{van Leeuwen}(2007)}]{VANLEEUWEN07}
{van Leeuwen}, F. 2007, \aap, 474, 653

\bibitem[{{Weber} \& {Strassmeier}(2011)}]{WEBER11}
{Weber}, M., \& {Strassmeier}, K.~G. 2011, \aap, 531, A89

\bibitem[{{Welsh} {et~al.}(2015){Welsh}, {Orosz}, {Short}, {Cochran}, {Endl},
  {Brugamyer}, {Haghighipour}, {Buchhave}, {Doyle}, {Fabrycky}, {Hinse},
  {Kane}, {Kostov}, {Mazeh}, {Mills}, {M{\"u}ller}, {Quarles}, {Quinn},
  {Ragozzine}, {Shporer}, {Steffen}, {Tal-Or}, {Torres}, {Windmiller}, \&
  {Borucki}}]{WELSH15}
{Welsh}, W.~F., {Orosz}, J.~A., {Short}, D.~R., {et~al.} 2015, \apj, 809, 26

\bibitem[{{Wizinowich} {et~al.}(2000){Wizinowich}, {Acton}, {Shelton},
  {Stomski}, {Gathright}, {Ho}, {Lupton}, {Tsubota}, {Lai}, {Max}, {Brase},
  {An}, {Avicola}, {Olivier}, {Gavel}, {Macintosh}, {Ghez}, \&
  {Larkin}}]{WIZINOWICH00}
{Wizinowich}, P., {Acton}, D.~S., {Shelton}, C., {et~al.} 2000, \pasp, 112, 315

\bibitem[{{W{\"o}llert} {et~al.}(2014){W{\"o}llert}, {Brandner}, {Reffert},
  {Schlieder}, {Mohler-Fischer}, {K{\"o}hler}, \& {Henning}}]{WOLLERT14}
{W{\"o}llert}, M., {Brandner}, W., {Reffert}, S., {et~al.} 2014, \aap, 564, A10

\bibitem[{{Wright} {et~al.}(2010){Wright}, {Eisenhardt}, {Mainzer}, {Ressler},
  {Cutri}, {Jarrett}, {Kirkpatrick}, {Padgett}, {McMillan}, {Skrutskie},
  {Stanford}, {Cohen}, {Walker}, {Mather}, {Leisawitz}, {Gautier}, {McLean},
  {Benford}, {Lonsdale}, {Blain}, {Mendez}, {Irace}, {Duval}, {Liu}, {Royer},
  {Heinrichsen}, {Howard}, {Shannon}, {Kendall}, {Walsh}, {Larsen}, {Cardon},
  {Schick}, {Schwalm}, {Abid}, {Fabinsky}, {Naes}, \& {Tsai}}]{WRIGHT10}
{Wright}, E.~L., {Eisenhardt}, P.~R.~M., {Mainzer}, A.~K., {et~al.} 2010, \aj,
  140, 1868

\bibitem[{{Yoss}(1961)}]{YOSS61}
{Yoss}, K.~M. 1961, \apj, 134, 809

\end{thebibliography}

\end{document}